\documentclass[11pt]{article}

\usepackage[T1]{fontenc}
\usepackage[utf8]{inputenc}
\usepackage{textcomp}
\usepackage{tgtermes}
\usepackage[version=3]{mhchem}
\usepackage{authblk}
\usepackage[colorlinks=false]{hyperref}
\usepackage[capitalize]{cleveref}
\usepackage{booktabs}
\usepackage{longtable}
\usepackage{amssymb}
\usepackage{lineno}
\usepackage{setspace}
\usepackage[dvipsnames]{xcolor}
\usepackage{geometry}
\usepackage{pdflscape}
\usepackage{upgreek}

\usepackage[title]{appendix}
\usepackage{natbib}
\bibpunct[]{(}{)}{;}{a}{,}{,}

\usepackage{bibunits}

\usepackage[dvipdfmx]{graphicx}

\graphicspath{{Fig/}}

\usepackage[finalnew]{trackchanges}

\addeditor{TY}
\addeditor{WFM}
\addeditor{wfm}

\title{\textbf{Variable refractory lithophile element compositions of planetary building blocks: insights from components of enstatite chondrites}}
\author[1]{Takashi Yoshizaki\thanks{Corresponding author. E-mail: \href{mailto:tky@dc.tohoku.ac.jp}{tky@dc.tohoku.ac.jp}}}
\author[2]{Richard D. Ash}
\author[2]{Marc D. Lipella\thanks{Deceased.}}
\author[3]{Tetsuya Yokoyama}
\author[1,2,4]{William F. McDonough}
\affil[1]{Department of Earth Science, Graduate School of Science, Tohoku University, Sendai, Miyagi 980-8578, Japan}
\affil[2]{Department of Geology, University of Maryland, College Park, MD 20742, USA}
\affil[3]{Department of Earth and Planetary Sciences, Tokyo Institute of Technology, Ookayama, Tokyo 152-8851, Japan}
\affil[4]{Research Center of Neutrino Sciences, Tohoku University, Sendai, Miyagi 980-8578, Japan}

\begin{document}

\doublespacing

\maketitle

\begin{bibunit}[elsarticle-harv]

\section*{Abstract}
\label{abstract}

Chondrites are sediments of materials left over from the earliest stage of the solar system history. Based on their undifferentiated nature and less fractionated chemical compositions, chondrites are widely considered to represent the unprocessed building blocks of the terrestrial planets and their embryos. \change{Compositional models of the planets}{Models of chemical composition of the terrestrial planets} generally find chondritic relative abundances of refractory lithophile elements (RLE) in the bulk \change{planets}{bodies} ("constant RLE ratio rule"), based on limited variations of RLE ratios among chondritic meteorites and the solar photosphere. Here, we show that ratios of RLE, such as Nb/Ta, Zr/Hf, Sm/Nd and Al/Ti, are fractionated \add{from the solar value} in chondrules from enstatite chondrites (EC). \remove{, which provides limitations on the use of the constant RLE ratio rule in the compositional modeling of planets} The fractionated RLE \change{compositions}{ratios} of individual EC chondrules document \add{different chalcophile affinities of RLE under highly reducing environments and} a separation of RLE-bearing sulfides from silicates before and/or during chondrule formation. \remove{ and different chalcophile affinities of RLE under highly reducing environments} \change{However, the solar-like RLE ratios of the bulk EC indicate a negligible widespread physical sorting of silicates and sulfides before and during the accretion of EC parent bodies.}{In contrast, the bulk EC have solar-like RLE ratios, indicating that a physical sorting of silicates and sulfides was negligible before and during the accretion of EC parent bodies.} Likewise, if the Earth's accretion were dominated by EC-like materials, as supported by multiple isotope systematics, the physical sorting of silicates and sulfides \add{in the accretionary disk} should not have occurred \change{in}{among} the Earth's building blocks.  Alternatively, the Earth's precursors might have been high-temperature nebular materials that condensed before \add{the RLE fractionation due to} precipitation of the RLE-bearing sulfides. A lack of Ti depletion in the bulk silicate Earth, combined with similar silicate-sulfide and rutile-melt partitioning behaviors of Nb and Ti, prefers a moderately siderophile behavior of Nb as the origin of the accessible Earth's Nb depletion. Highly reduced planets that have experienced selective removal or accretion of silicates or metal/sulfide phases, such as Mercury, might have fractionated, non-solar bulk RLE ratios. \bigskip

\noindent Keywords: planets, chondrites, Earth, refractory lithophile elements, chemical fractionation


\section{Introduction}
\label{sec:intro}

The geochemical classification of elements is established based on their partitioning behaviors during condensation into a silicate (lithophile), metal (siderophile) or sulfide (chalcophile) phases \citep[e.g.,][]{lodders2003solar}. \remove{with a cosmochemical reference frame of their volatilities} \add{Elements are also classified cosmochemically based on their volatilities.} There are 36 refractory elements, and most of them are generally considered to be lithophile (e.g., Be, Al, Ca, Ti, Sc, Sr, Y, Zr, Nb, Ba, rare earth elements (REE), Hf, Ta, Th and U), while V, Mo, and W are moderately siderophile/chalcophile, and Ru, Rh, Re, Os, Ir, and Pt are highly siderophile. Chondritic meteorites (chondrites) are undifferentiated solar system materials that are widely considered as planetary building blocks. The relative abundances of the most refractory lithophile elements (RLE) are nearly constant ($ \lesssim $10\%) for the solar photosphere and among chondritic meteorites   \citep{larimer1988refractory,wasson1988compositions,bouvier2008lu}. Importantly, compositional models of Earth \citep{allegre1995chemical,mcdonough1995composition,palme2014cosmochemical} and Mars \citep{wanke1994chemistry,yoshizaki2019mars_long}  commonly find or assume chondritic RLE ratios (e.g., Ca/Al, Sm/Nd, Lu/Hf, Th/U) for these planets ("constant RLE ratio rule"). \bigskip

However, recent high-precision measurements of chondritic and terrestrial samples revealed \change{meaningful}{detectable} variations of not only REE ratios \citep{dauphas2015thulium,barrat2016evidence}, but also geochemical twins' ratios such as Y/Ho \citep{pack2007geo}, Zr/Hf \citep{patzer2010zirconium}, and Nb/Ta \citep{munker2003evolution} among these solar system materials. For example, the observation of lower Nb/Ta \remove{value} of the accessible Earth as compared to CI chondrites, which show the closest match to the solar photosphere composition for many elements \citep[e.g.,][]{lodders2020solar}, raised considerable discussions on their relative behaviors. Some authors proposed that Nb and Ta are exclusively lithophile and hosted in a eclogitic reservoir hidden in the mantle \citep{sun1989chemical,mcdonough1991partial,kamber2000role,rudnick2000rutile,nebel2010deep}, whereas others have suggested that Nb and less so Ta behaved as moderately siderophile \citep{wade2001earth,munker2003evolution,corgne2008metal,mann2009evidence,cartier2014redox} or chalcophile \citep{munker2017silicate} elements and have been partially sequestered into the metallic core. \bigskip

The Earth's specific building blocks remain unknown. Among chondritic meteorites, enstatite chondrites (EC) show the closest matches to the Earth's mantle in multiple isotopic systematics \citep{javoy2010chemical,warren2011stable,dauphas2017isotopic,boyet2018enstatite}. On the other hand, EC are characterized by lower Mg/Si and RLE/Mg ratios and higher volatile abundances as compared to Earth \citep{palme1988moderately,mcdonough1995composition}. EC \change{is a}{are} rare type of chondrite \add{in our collection} that record the most reducing conditions among chondritic meteorites
\citep[\cref{fig:UC_diagram};][]{keil1968mineralogical,larimer1968experimental}. The chemical and isotopic composition of meteorites and Earth, combined with partitioning behaviors of elements, indicate that an early stage of the Earth's formation was dominated by an accretion of EC-like, highly reduced materials, followed by an accretion of less reduced ones \citep[e.g.,][]{corgne2008metal,mann2009evidence,schonbachler2010heterogeneous,rubie2011heterogeneous,rubie2015accretion}. It is known that multiple RLE show non-lithophile behaviors under highly reduced conditions as recorded in EC \citep[e.g.,][]{barrat2014lithophile,munker2017silicate}, but contributions of such characteristics of RLE to the compositions of Earth and its building blocks remain poorly understood. \bigskip

Here, we report RLE composition of chondrules, one of the major RLE carriers in chondritic meteorites \citep[e.g.,][]{alexander2005re,alexander2019quantitative_CC,alexander2019quantitative_NC} and potential dominant source materials of the terrestrial mantle \citep[e.g.,][]{hewins1996nebular,johansen2015growth,levison2015growing,amsellem2017testing,yoshizaki2020earth}, and sulfides from primitive EC samples, to reveal non-lithophile behavior of RLE in highly reducing environments. Using these and published data from different types of chondrites, combined with some new RLE data for refractory inclusions from carbonaceous chondrites (CC), we examine a compositional variability in RLE ratios within the accretionary disk, behavior of elements under variable redox conditions, and the chondritic reference frame for bulk planetary compositions. \bigskip

\section{Samples and methods}

\subsection{Samples}

In order to reduce potential modifications of elemental abundances by parent body metamorphism and terrestrial weathering, we chose unequilibrated (type 3) EH, EL and CV chondrites with limited terrestrial weathering as samples for this study (\cref{tab:sample_list}). Analyses were conducted on 92 chondrules, 40 troilites (FeS), 9 oldhamites (CaS) and 5 niningerites (MgS) from Alan Hills (ALH) A77295 (EH3), ALH 84170 (EH3), ALH 84206 (EH3), ALH 85119 (EL3) and MacAlpine Hills (MAC) 88136 (EL3). We also studied two Ca, Al-rich inclusions (CAI) from Allende (CV3), and one CAI and six amoeboid olivine aggregates (AOA) from Roberts Massif (RBT) 04143 (CV3). The petrology and mineralogy of these meteorites have been well characterized in previous studies \citep[e.g.,][]{johnson1995relative,rubin1997shock,benoit2002thermoluminescence,lin2002comparative,grossman2005onset,gannoun2011ree,quirico2011reappraisal,ishida2012diverse,yoshizaki2019nebular,bonal2020water}. \bigskip

\subsection{Analytical methods}

Mineralogy, petrology and major element compositions of polished sections of the samples were analyzed using scanning electron microscopes (SEM) and electron probe microanalyzers (EPMA) at Tohoku University and University of Maryland, using procedures after \citet[][; Section B.1]{yoshizaki2019nebular}. Trace element compositions of chondritic components were determined using a New Wave frequency-quintupled Nd:YAG laser system coupled to a Thermo Finnigan Element2 single-collector ICP-MS at University of Maryland, following procedures of \citet{lehner2014eh3}. Operating conditions are summarized in \cref{tab:LA-ICP-MS_cnd}. Measurements were carried out in low-mass resolution mode (M/$\Delta$M = 300), with 15--200 $\upmu$m laser spot size and a fluence of $\sim$2--3 J/cm$^{2}$. \change{In all circumstances, the spot size was set to be smaller than the chondrule so that the laser sampled only the interiors of chondrules.}{In all circumstances, the spot size was set to be smaller than the chondrule so that the laser sampled only the polymineralic interiors of chondrules.} The beam was focused onto a sample placed in a 3 cm$ ^3 $ ablation cell, which was continuously flushed to the plasma source of the mass spectrometer with a He gas flow of $ \sim $1 L/min. The mass spectrometer was tuned to maximize signal (based on \ce{^{43}Ca} and \ce{^{232}Th} spectra) and minimize oxide production (UO/U $<$ 1.2\%) to maximize sensitivity and reduce \change{isobaric}{polyatomic} interference. A dwell time of 5--15 milliseconds was used depending on element concentrations. After the LA-ICP-MS measurements, the sampled areas were investigated using SEM, to make sure that there was no contamination due to beam overlapping onto neighboring phases (e.g., matrix). \bigskip

In LA-ICP-MS measurement of chondrules and refractory inclusions, National Institute of Standards and Technology (NIST) 610 glass \citep{jochum2011determination} was used as an external standard material. Internal standards were \ce{^{29}Si} for chondrules, and \ce{^{29}Si}, \ce{^{43}Ca} or \ce{^{47}Ti} for refractory inclusions, which were determined as a polymineralic bulk composition using EPMA. For sulfides, the group IIA iron meteorite Filomena (for siderophile elements) and the NIST 610 glass (for lithophile elements) were used as reference materials, using data from \citet{walker2008modeling} and the GeoReM database \citep{jochum2011determination} as working values, respectively. Count rates were normalized using \ce{^{57}Fe} (for Filomena) and \ce{^{63}Cu} (for NIST 610) as internal standards, which were determined using EPMA or SEM/EDS. The standard materials were measured at least twice at the start of individual analytical run that is composed of 16 analyses of unknown samples. All obtained data were processed using the LAMTRACE software \citep{van2001data}. Data with a potential beam overlapping onto neighboring phases were rejected in this procedure. The long-term reproducibilities of isotope ratios for the external standards were better than 5\% for nearly all of the measured elements, except for Ni and Cu in Filomena (Table B.1). We also routinely measured the BHVO-2G glass as a secondary standard material in order to monitor accuracy of the analysis, which well reproduced the preferred values \citep{jochum2008reference} within $ < $5\%. \add{We report mean $ \pm $ 2 standard errors of the mean (2$ \sigma_{\mathrm{m}} $)) of RLE ratios of EC chondrules, EC sulfides, and CC refractory inclusions, as a measure of their average RLE ratios.}  \bigskip

\section{Results}
\label{sec:results}

\subsection{Chondrules from unequilibrated enstatite chondrites}
\label{sec:results_chondrules}

Various types of chondrules are identified and measured in this study. More than 70\% of them are porphyritic pyroxene (PP) chondrules, with less abundant ($ \sim $15\%) radial pyroxene (RP),  olivine-bearing (porphyritic olivine pyroxene (POP) or porphyritic olivine (PO) type; $ \sim $10\%), and other rare types (\cref{fig:chondrule_volatility_PP,fig:chondrule_volatility_others}). These chondrules are mostly 200--600 $ \upmu $m in diameter, with some reaching up to 1 mm. The observed abundance of chondrule types and their size distributions are consistent with previous reports \citep[e.g.,][]{jones2012petrographic,weisberg2012unequilibrated,jacquet2018chondrules}. \bigskip

PP chondrules are mainly composed of Fe-poor ($ < $2 wt\%) enstatite and mesostasis, with minor \change{oxides}{silicates} such as olivine and silica, as reported by previous studies \citep[e.g.,][]{weisberg2012unequilibrated,jacquet2018chondrules}. They sometimes contain small amounts of opaque phases, such as Si-bearing Fe-Ni metal, troilite, oldhamite, niningerite (only in EH3) and alabandite (MnS; only in EL3). Although each PP chondrule shows variable lithophile element abundances, with absolute RLE enrichment factors of 0.5--2 $ \times $ CI (\cref{fig:chondrule_volatility_PP}A--E), their average compositions are similar in each sample (\cref{fig:chondrule_volatility_PP}F). On average, PP chondrules are characterized by depletions of REE, Y, high-field strength elements (HFSE: Ti, Zr, Hf, Nb, Ta), Cr, Mn and K compared to Al, Mg and Si. REE and HFSE are also fractionated within these groups; Eu and Yb are depleted compared to other REE, and Zr and Nb show stronger depletions than their geochemical twins, Hf and Ta, respectively. \change{Elevated REE and Y abundances in one PP chondrules from MAC 88136 (C14) might reflect its oldhamite-bearing mineralogy.}{One PP chondrule from MAC 88136 (C14) has numerous tiny grains of oldhamite in its glassy area, which might have contributed its elevated bulk REE and Y abundances.} \bigskip

Compared to PP chondrules, RP chondrules show lower RLE abundance (0.2--1 $ \times $ CI; \cref{fig:chondrule_volatility_others}A,B), whereas PO and POP chondrules have higher RLE abundances (0.8--4 $ \times $ CI; \cref{fig:chondrule_volatility_others}C). One cryptocrystalline chondrule shows flat RLE pattern with a strong negative Eu anomaly and Al, Mg and Si enrichments (\cref{fig:chondrule_volatility_others}D). In general, RP, olivine-bearing, and cryptocrystalline chondrules are enriched in Al, Mg and Si compared to most RLE and Cr, Mn and K, as observed in PP chondrules. \add{We note that these variations are potentially affected by sectioning effects (see Section B.2).} A \change{Ca-rich}{Ca-px-rich} chondrule in ALH 84206 shows a flat, elevated RLE pattern ($ \sim $ 6 $ \times $ CI), with clear depletions of Ce, Ba, Cr, Mn and K (\cref{fig:chondrule_volatility_others}D). In addition, EC chondrules show Mg/Si (0.84 $ \pm $ 0.01, 2$ \sigma_{\mathrm{m}} $)\add{, which is} lower than those of CI 
\citep[0.89;][]{lodders2020solar} and the bulk silicate Earth (BSE) \citep[1.1;][]{mcdonough1995composition,palme2014cosmochemical}. \bigskip


EC chondrules show highly variable, fractionated ratios for multiple RLE pairs (Figs.~B.3 and 4). Their RLE ratios are not clearly related to their petrological types (e.g., PP, PO, POP). The most significant ones are ratios of Nb to other RLE, with average Nb/Ta (13.4 $ \pm $ 1.6, \change{2 standard errors of the mean (2$ \sigma_{\mathrm{m}} $)}{2$ \sigma_{\mathrm{m}} $)}) and Nb/La (0.691 $ \pm $ 0.131, 2$ \sigma_{\mathrm{m}} $) being $ \sim $30\% and $ \sim $40\% lower than the CI ratios \citep[$ \sim $19 and $ \sim $1.1;][]{munker2003evolution,barrat2012geochemistry,braukmuller2018chemical,lodders2020solar}, respectively. EC chondrules, on average, also show lower Zr/Hf (30.0 $ \pm $ 1.5, 2$ \sigma_{\mathrm{m}} $), Sm/Nd (0.281 $ \pm $ 0.021, 2$ \sigma_{\mathrm{m}} $) and Ca/Al (0.848 $ \pm $ 0.061, 2$ \sigma_{\mathrm{m}} $) values, and elevated Al/Ti (33.3 $ \pm $ 3.1, 2$ \sigma_{\mathrm{m}} $) as compared to the CI values (i.e., $ - $12\%, $ - $15\%, $ - $20\% and $ + $75\% relative to CI, respectively). On the other hand, an average ratio of Y and its geochemical twin Ho in EC chondrules (25.8 $ \pm $ 0.7, 2$ \sigma_{\mathrm{m}} $) overlaps the CI composition. \bigskip

It should be noted that the bulk composition of individual chondrules derived by surface analytical methods (i.e., broadened or scanning beam measurements on thin or thick sections) can be less accurate than those determined by a whole-rock measurement of mechanically separated chondrules (see Section B.2). Our data for Nb/La, Zr/Hf, Al/Ti, and Ca/Al in EC chondrules are consistent with EH4 chondrule compositions from \citet{gerber2012chondrule}, who measured elemental abundances of solutions of mechanically isolated individual chondrules using ICP-MS and ICP-OES techniques. Slightly elevated Y/Ho and Sm/Nd ratios of EH4 chondrules from \citet{gerber2012chondrule} might reflect distinct behavior of these elemental twins during thermal processing in the EC parent bodies \citep{barrat2014lithophile}. We consider this consistency as reflecting representative sampling of the bulk EC chondrule compositions in our measurements. Furthermore, our bulk EC chondrule data are generally consistent with results from previous studies which used surface analytical methods \citep{grossman1985chondrules,schneider2002properties,varela2015nonporphyritic}. Further discussions on the comparison of data obtained in this and previous studies are provided in Section B.2. \bigskip

\subsection{Sulfides from unequilibrated enstatite chondrites}
\label{lab:sec_sulfides}

Both EH3 and EL3 chondrites contain abundant opaque nodules with variable mineralogies. They are commonly composed of Si-bearing Fe-Ni metal and troilite, with minor occurrence of exotic sulfides including oldhamite, daubr\'{e}elite (\ce{Fe^{2+}Cr^{3+}_2S4}), niningerite (in EH3), alabandite (in EL3), K-bearing sulfides, and phosphides. We report only trace element abundances of EC sulfides which occur outside of EC chondrules, as those found within chondrules are tiny (mostly $ < $10 $ \upmu $m), making analysis by LA-ICP-MS impractical. \bigskip

Troilites from the primitive EC contain RLE such as Ti (2--7 $ \times $ CI), Nb (mostly 0.1--5 $ \times $ CI) and Zr (up to 0.2 $ \times $ CI), along with nominally chalcophile Cr, Mn and V (\cref{fig:FeS_RLE}; see also Section B.3). In contrast, Ta and Hf, which are geochemical twins of Nb and Zr, respectively, were not detected in the EC troilites. Oldhamites in EC are enriched in multiple RLE, including Zr, Hf, Y, REE, Th and U (Fig.~B.4). Most oldhamites show nearly flat REE pattern at 10--100 times higher concentrations than CI, with slightly positive Eu and Yb anomalies in those from EH. They show Zr/Hf (40--600) higher than CI, and variable Sm/Nd (0.34 $ \pm $ 0.11, 2$ \sigma_{\mathrm{m}} $), Y/Ho (31 $ \pm $ 7, 2$ \sigma_{\mathrm{m}} $) and Th/U (4.5 $ \pm $ 2.5, 2$ \sigma_{\mathrm{m}} $) values, which are generally consistent with data reported in \citet{gannoun2011ree}. Niningerites in EH are enriched in Sc (2--12 $ \times $ CI) and Zr (0.2--16 $ \times $ CI) (Table E.3). Aluminum was not detected in nearly all of sulfide measurements, indicating that it retains its lithophile behavior under highly reducing conditions. \bigskip

\subsection{Refractory inclusions from unequilibrated carbonaceous chondrites}
\label{sec:results_refractory_inclusions}

CAI and AOA from CV chondrites show highly fractionated RLE compositions (Figs.~B.5 and B.6; see also Section B.4), reflecting the volatility-driven fractionation of RLE during condensation of these inclusions \citep[e.g.,][]{boynton1975fractionation}. Their RLE compositions are generally consistent with previous studies \citep[e.g.,][ and references therein]{boynton1975fractionation,mason1977geochemical,kornacki1986abundance,stracke2012refractory,patzer2018chondritic}. CAI and spinel-rich AOA show low Nb/Ta ratios (3--10), whereas olivine-rich AOA show higher values (18--37). The CV CAI and AOA also show fractionated Zr/Hf (mostly 33--55), Y/Ho (17--38), and Sm/Nd (0.12--0.54) values. \bigskip

\section{Discussion}

\subsection{RLE fractionation under reduced conditions}
\label{sec:RLE_fractionation}

Ratios of RLE in individual chondrules from EC, OC, and CC can provide useful insights into distribution and fractionation processes of RLE in the accretionary disk. Although chondrules show considerable scatter in RLE ratios (\cref{fig:chondrule_RLEratio}), their mean compositions provide a measure of distinction between each chondrite class (EC, OC, and CC), reflecting limited mixing of distinct chondrule reservoirs \citep[e.g.,][]{clayton2003oxygen_tog,hezel2018spatial}. Here we propose that RLE fractionation processes recorded in highly reduced EC chondrules are distinct from those in less reduced OC and CC chondrules. \bigskip

It is well recognized and demonstrated that an incomplete high-temperature condensation from the gas of solar composition can fractionate RLE, as recorded in CAI and AOA \citep[Fig.~B.5; e.g.,][]{boynton1975fractionation,kornacki1986abundance,ruzicka2012amoeboid}. 
Nb/Ta ratios of all studied CAI and Ca, Al-rich AOA range from 3 to 10 (Fig.~B.6), which are clearly lower than the CI value. In contrast, olivine-rich, less refractory AOA do not show such low Nb/Ta ratios. The low Nb/Ta ratios observed in CAI and refractory-rich AOA might reflect Nb-Ta fractionation during formation of high-temperature nebular condensates, due to slightly different condensation temperatures of these geochemical twins (1559 K and 1573 K for Nb and Ta, respectively) \citep{lodders2003solar,munker2003evolution}. Similarly, a slightly higher 50\% condensation temperatures of Zr and Y compared to their geochemical twins Hf and Ho, respectively, can produce condensates with non-CI Zr/Hf and Y/Ho ratios \citep[Figs.~B.5 and B.6B,E;][]{el2002efremovka,pack2007geo,patzer2010zirconium,patzer2018chondritic,stracke2012refractory}. The non-CI Nb/Ta, Zr/Hf and Y/Ho values in bulk OC and CC chondrites have been attributed to additions of the refractory materials with fractionated RLE compositions into OC and CC asteroids \citep{munker2003evolution,pack2007geo,patzer2010zirconium,stracke2012refractory}. Likewise, these additions might also be responsible for chondrules possessing similar fractionated Nb/Ta and Zr/Hf ratios \citep[e.g.,][]{misawa1988demonstration,misawa1988highly,pack2004chondrules,patzer2018chondritic}. In contrast, the CI-like average Y/Ho in OC and CC chondrules indicate a similar behavior of these geochemical twins during formation of these chondrules and their precursors \citep{pack2007geo}. \bigskip

Compared to OC and CC chondrules, EC chondrules show more distinct differences in key RLE ratios (e.g., Nb/Ta, Nb/La, Zr/Hf, Sm/Nd, Al/Ti, Ca/Al; Figs.~\ref{fig:chondrule_RLEratio} and B.3). Importantly, most RLE are depleted relative to Al in EC chondrules (e.g., super-CI Al/Ti and sub-CI Ca/Al); this feature is distinct from CI-like RLE/Al values in OC and CC chondrules (\cref{fig:chondrule_volatility_PP,fig:chondrule_volatility_others,fig:chondrule_RLEratio}). These differences indicate that RLE fractionation processes recorded in EC chondrules differ from those of OC and CC chondrules \citep[i.e., volatility-driven fractionation;][]{misawa1988demonstration,misawa1988highly,pack2004chondrules,patzer2018chondritic}. \bigskip

The fractionated RLE ratios in EC chondrules might indicate depletion of moderately chalcophile elements under highly reducing conditions \citep[e.g., Ca, REE, Ti, Nb, Zr; Figs.~\ref{fig:FeS_RLE} and B.4;][]{gannoun2011ree,barrat2014lithophile,lehner2014eh3} due to separation of sulfides from silicates. The high Al/Ti and low Nb/Ta and Zr/Hf in EC chondrules are consistent with separation of Ti, Nb, Zr-bearing troilites from a precursor with CI composition. The sub-CI Ca/Al and REE/Al in EC chondrules are in harmony with a removal of REE-bearing oldhamite. The separation of oldhamites can also produce sub-CI Sm/Nd and negative Eu and Yb anomalies of EC chondrules (\cref{fig:chondrule_volatility_PP,fig:chondrule_volatility_others,fig:chondrule_RLEratio}), given high Sm/Nd and positive Eu and Yb anomalies in oldhamites \citep[Fig.~B.4;][]{gannoun2011ree,jacquet2015formation}. In addition, smaller depletions of Hf and Ta as compared to their geochemical twins Zr and Nb, respectively (\cref{fig:chondrule_volatility_PP,fig:chondrule_volatility_others}), indicate the former's more lithophile behavior \citep{barrat2014lithophile,munker2017silicate}. Furthermore, The CI-like Y/Ho values of EC chondrules are consistent with limited fractionation of Y/Ho in EC oldhamites (31.3 $ \pm $ 6.7, 2$ \sigma_{\mathrm{m}} $; Table A.3) from the CI ratio \citep[$ \sim $26;][]{pack2007geo,barrat2012geochemistry,lodders2020solar}. The stronger depletion of the troilite-loving elements (e.g., Ti, Nb) compared to the oldhamite-loving ones' (e.g., REE, Y) in EC chondrules documents a greater contribution of troilite separation to their fractionated RLE compositions. This observation is consistent with a simple mass-balance consideration of modal abundances of sulfides in type 3 EC \citep{weisberg2012unequilibrated} and their average RLE concentrations \citep[this study;][]{gannoun2011ree} that indicate the separation of RLE-bearing troilites produces $ > $2 times larger RLE fractionation in residual silicates than the removal of oldhamite. \bigskip

Chondrules from all types of chondrites are depleted in normally siderophile and chalcophile elements, due to a separation of metal and sulfide components from silicates before and/or during chondrule formation  \citep{osborn1973elemental,gooding1980elemental,grossman1985chondrules,palme2014siderophile}. For example, it is suggested that metals and sulfides can be expelled from silicate melt droplets during the chondrule formation as immiscible melts \citep{rambaldi1981metal,rambaldi1984metal,grossman1985origin,zanda1994origin,mccoy1999partial,connolly2001formation,uesugi2008kinetic}. In addition, if metal/sulfide and silicate occurred separately in the protoplanetary disk as a result of the chondrule formation and/or nebular condensation, they could have been physically sorted due to their distinct density, size, magnetism, and/or thermal conductivity \citep{kuebler1999sizes,wurm2013photophoretic,kruss2020composition,palme2014siderophile}. These separation processes can also contribute to the chalcophility-dependent fractionation of elements in EC chondrules. \bigskip

The origin of metal-sulfide nodules in EC remains controversial. One view is that they are highly reduced nebular condensates formed and processed separately from silicates before planetesimal formation \citep{larimer1975effect,larimer1979role,larimer1987trace,el1988qingzhen,kimura1988origin,lin2002comparative,gannoun2011ree,el2017origin}, whereas others propose that they were ejected from chondrule-forming melt droplets as immiscible liquids  \citep{hsu1998geochemical,horstmann2014clues,lehner2014eh3,ebel2015complementary,jacquet2015formation,piani2016magmatic} or combination of these processes \citep{lehner2010origin}. The post-accretionary impact origin of these nodules is inconsistent with a lack of petrological and chemical features of in-situ melting  \citep{horstmann2014clues,gannoun2011ree,el2017origin}. \bigskip

\begin{figure}[h]
\centering
\includegraphics[width=0.7\linewidth]{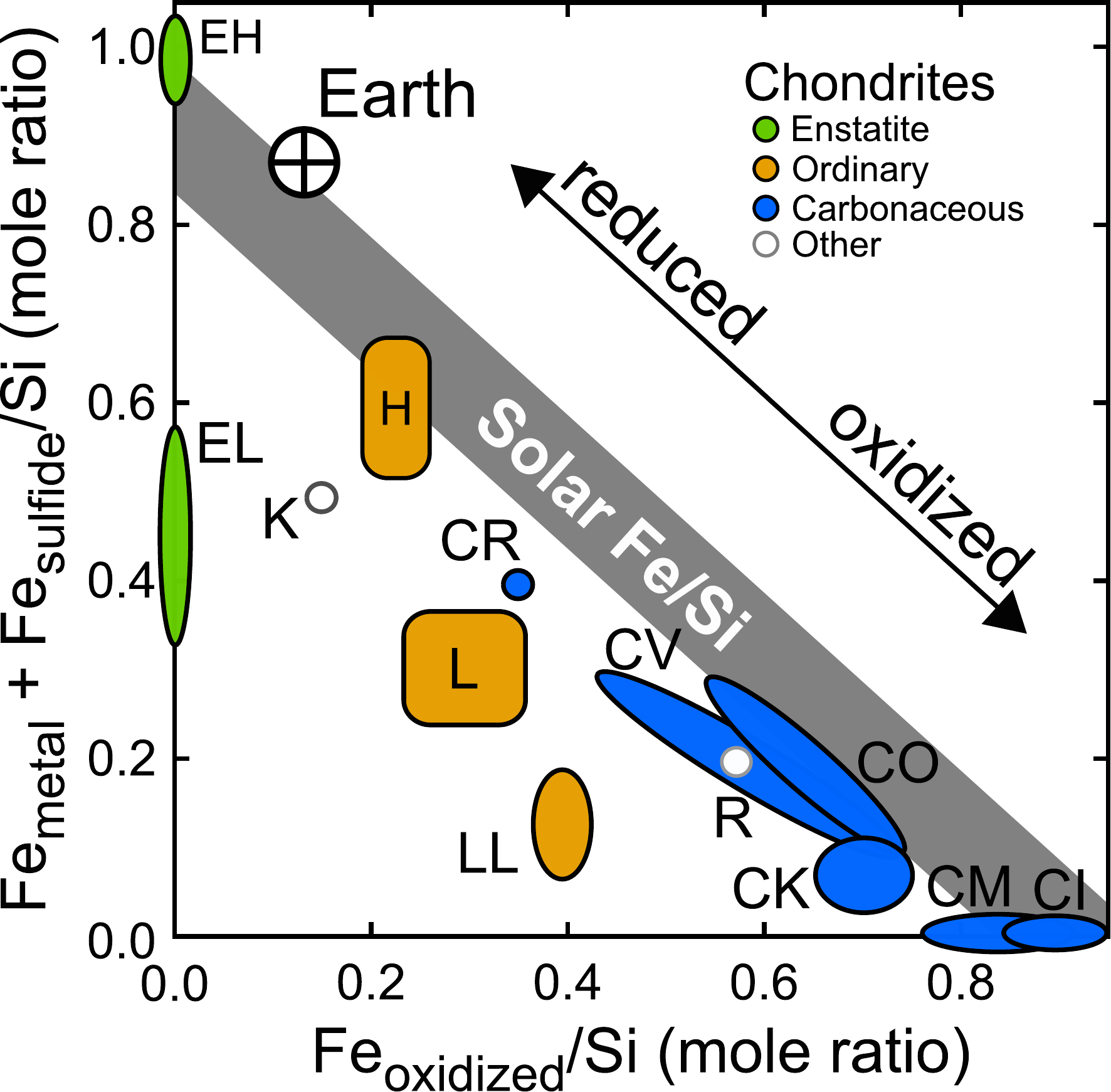}
\caption{The Urey-Craig diagram showing variation in oxidation state of iron relative to silicon among chondrite groups \citep[after][]{urey1953composition}. The plot of Earth is based on \citet{mcdonough2014compositional}. Metal-rich CB and CH chondrites are not included in the plot because their origin \citep[condensation from a gas-melt plume formed by a planetary-scale collision; e.g.,][]{krot2012isotopically} is distinct from that of major chondrite groups.} 
\label{fig:UC_diagram}
\end{figure}

\begin{figure}[h]
\centering
\includegraphics[width=1.0\linewidth]{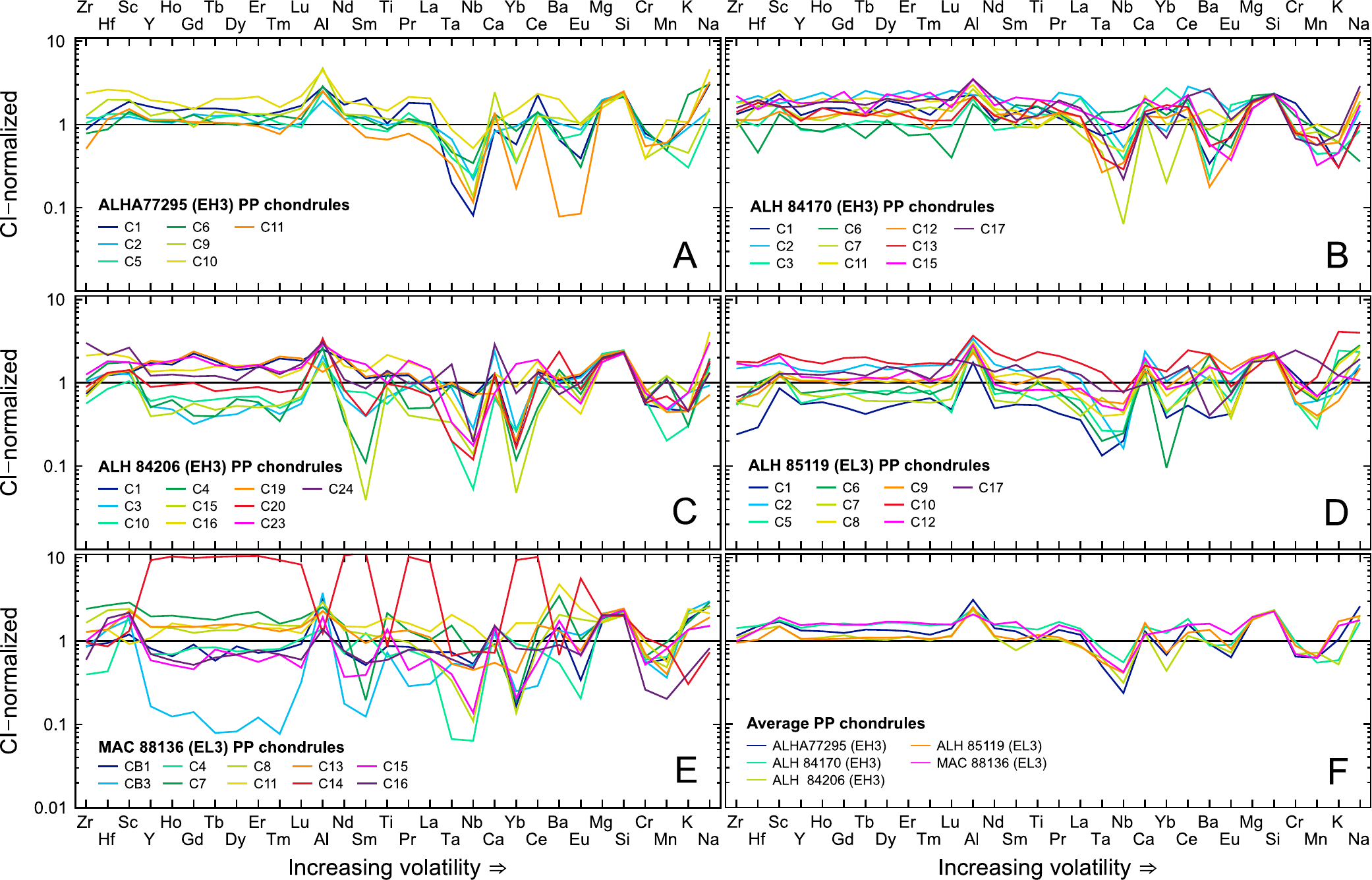}
\caption{(A--E) CI-normalized \citep{lodders2020solar} abundances of lithophile elements in selected porphyritic pyroxene (PP) chondrules from unequilibrated enstatite chondrites, and (F) average composition of PP chondrules in each sample.}
\label{fig:chondrule_volatility_PP}
\end{figure}

\begin{figure}[h]
\centering
\includegraphics[width=1.0\linewidth]{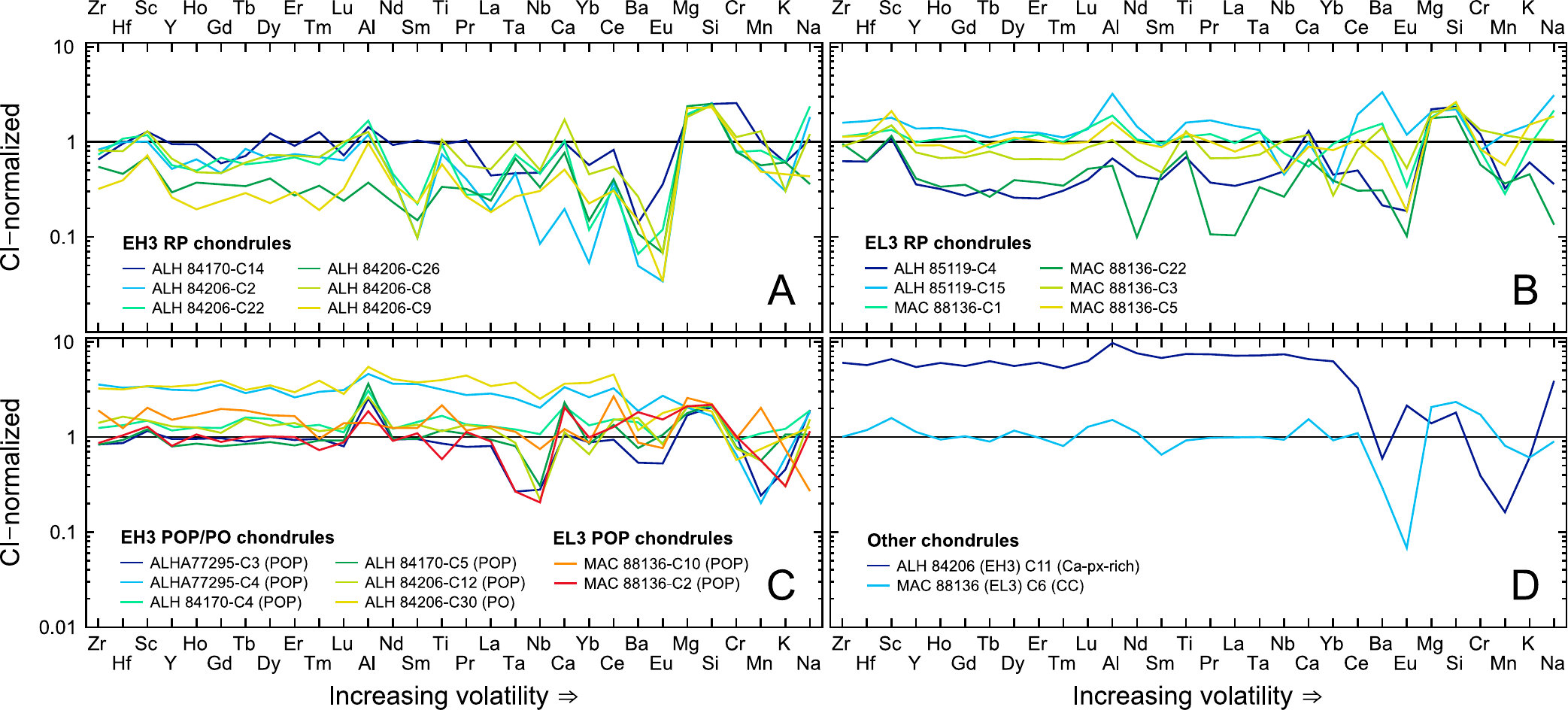}
\caption{CI-normalized \citep{lodders2020solar} abundances of lithophile elements in non-PP chondrules from unequilibrated enstatite chondrites. Abbreviations: CC--cryptocrystalline; PO--porphyritic olivine; POP--porphyritic olivine pyroxene; RP--radial pyroxene.}
\label{fig:chondrule_volatility_others}
\end{figure}

The formation of metal-sulfide nodules and their separation from silicates can be critical processes in establishing Fe/Si ratio of bulk bodies with highly reduced oxidation states. The lower bulk Fe/Si value of EL chondrites as compared to EH (\cref{fig:UC_diagram}) can be partly attributed to less abundant occurrence of troilite in EL \citep[$ < $8 vol\% and $ > $10 vol\% in primitive EL and EH, respectively;][]{weisberg2012unequilibrated}. If the low troilite abundance in EL reflect the ejection of sulfides from chondrule-forming silicate melts and their limited accretion to the EL parent bodies, the bulk EL should be depleted in moderately chalcophile RLE. However, neither \change{bulk EH3 and EL3 chondrites do not show}{bulk EH3 nor EL3 shows} fractionated RLE compositions \citep{barrat2014lithophile}. Even ratios of RLE with different chalcophile affinities (e.g., Nb/La, Zr/Hf, Sc/Hf, Sc/Nb) are CI-like in the bulk EC \citep{barrat2014lithophile}, which are distinct from  highly fractionated RLE compositions of their chondrules  (\cref{fig:chondrule_volatility_PP,fig:chondrule_volatility_others,fig:chondrule_RLEratio}). These observations suggest that the depletion of moderately chalcophile RLE in EC silicates (i.e., chondrules) are fully compensated by the occurrence of RLE-bearing sulfides in the bulk EC, and \change{silicate-sulfide separation during the chondrule formation}{removal of sulfides from chondrule-forming melts} did not contribute to the low troilite abundance in EL. Thus, the metal- and sulfide-depletion in EC chondrules and the CI-like RLE ratios of the bulk EC indicate that EC silicates and metal-sulfide nodules occurred \change{separately}{as separate objects} in the protoplanetary disk, but \change{they were not physically sorted}{none of them was preferentially removed from the reservoir} before and during the EC parent body accretion. The limited physical sorting of the EC silicates, sulfides, and metals indicates that their separation was a localized process in the protoplanetary disk, and EC parent bodies accreted quickly after the chondrule formation events. \add{The fractionated RLE ratios of EC components (Fig. 4) and solar-like RLE ratio of the bulk EC (Barrat et al., 2014) might support a compositional complementarity between chondrules and other components (e.g., Hezel et al., 2018).} \nocite{barrat2014lithophile,hezel2018composition} \bigskip

In order to further constrain the timing and mechanism of RLE fractionation recorded by EC chondrules and sulfides, especially those in EH, it demands an understanding of the redox-dependent changes of RLE volatilities. Although multiple equilibrium condensation calculations have been performed for highly reducing conditions \citep[e.g.,][]{larimer1975effect,larimer1979role,lodders1993lanthanide,wood1993mineral,sharp1995molecular,ebel2006condensation,ebel2011equilibrium}, relative volatilities of minor and trace RLE, especially those for troilite-loving ones (e.g., Ti, Nb), are poorly constrained. A combination of relative volatilities and chalcophile affinities of RLE under reducing conditions will provide important constraints on the RLE fractionation mechanism \change{during the formation of planetary building blocks}{prior to the accretion of EC parent bodies}. \bigskip


\subsection{Compositional \change{distinction}{difference} between EC chondrules and the silicate Earth}
\label{sec:no_reduced_sulfide}

The limited variation ($ < $10\%) in RLE ratios among chondritic meteorites and the solar photosphere have led planetary scientists to use the constant RLE ratio rule in their compositional models of planets (Section 1). \change{However, the fractionated RLE ratios of EC chondrules (Figs. 2 to 4) places limits on the use of this rule, especially when modeling composition of highly reduced bodies (e.g., Mercury; Section 4.4).}{In contrast, the fractionated RLE ratios of EC chondrules (Figs. 2 to 4) potentially places limits on the use of this rule, when modeling composition of highly reduced bodies (e.g., Mercury; Section 4.4).} In turn, these variations can be used as a new key to reveal the nature of planetary building blocks, together with the isotopic constraints \citep[e.g.,][]{javoy2010chemical,warren2011stable,dauphas2017isotopic}. \bigskip

Isotopic similarities of Earth and EC suggest EC-like materials as the Earth's main building block \citep[e.g.,][]{javoy2010chemical,warren2011stable,dauphas2017isotopic,boyet2018enstatite}. In this case, the Earth's mantle could have been dominated by EC chondrule-like materials, as they host nearly all of silicates in EC \citep[e.g.,][]{weisberg2012unequilibrated,krot2014classification}. EC chondrules and the Earth's mantle share Nb depletions, with Nb/Ta and Nb/La values being 20--30\% lower than the CI chondritic values \citep[\cref{fig:chondrule_RLEratio};][]{munker2003evolution}. In contrast, composition of terrestrial komatiites\add{, high-temperature ultramafic rocks mostly formed in the Archean,} is consistent with a CI-like Ti/RLE values for the Earth's primitive mantle, and thus there is no \change{"missing Ti problem"}{Ti depletion} in the BSE \citep[e.g.,][]{nesbitt1979komatiites,mcdonough1995composition,arndt2003komatiites}. The fractionated Al/Ti values of EC chondrules highlight their compositional distinction from the BSE. \bigskip

Experimental studies show that Nb and Ti have similar $ D^{\mathrm{sulfide/silicate}} $ values at wide $ P $, $ T $, and $f_{\ce{O2}}$ conditions, with the value exceeding 1 at highly reduced conditions \citep[$ \log f_{\ce{O2}} \leq \mathrm{IW} -  2;$, \cref{fig:D-values}A;][]{kiseeva2015effects,wood2015trace,namur2016sulfur,munker2017silicate,cartier2020no,steenstra2020geochemical}. Compositions of troilites and silicates from aubrites, highly reduced achondrites, support similar chalcophilities of Nb and Ti during differentiation of aubrite parent bodies \citep{van2012siderophile,munker2017silicate}. These observations suggest that a sulfide-silicate fractionation under highly reduced conditions produces not only Nb, but also Ti depletion in silicates at a similar degree\change{, which is not observed in the BSE.}{. However, unlike Nb, Ti is not depleted in the BSE (e.g., Nesbitt et al., 1979; McDonough and Sun, 1995; Arndt, 2003).} \nocite{nesbitt1979komatiites,mcdonough1995composition,arndt2003komatiites}  Thus, even if the EC chondrules are the Earth's main building block \citep{javoy2010chemical,warren2011stable,dauphas2017isotopic,boyet2018enstatite}, the accessible Earth's Nb depletion should not have resulted from a sulfide-silicate separation under highly reduced conditions. The negligible role of highly reduced sulfide in the present-day Earth's composition is also consistent with CI-like REE abundance of the BSE \citep[e.g.,][]{mcdonough1995composition,pack2007geo,bouvier2008lu,bouvier2016primitive,burkhardt2016nucleosynthetic} \citep[cf.][]{stracke2012refractory,dauphas2015thulium} and a lack of evidence for Th and U incorporation into the Earth's core \citep{wipperfurth2018earth}. \bigskip

The relative partitioning behaviors of Nb and Ti are also useful when considering a compositional evolution of the Earth's mantle. Sub-CI Nb/RLE and CI-like Ti/RLE ratios of komatiites and Archean basalts indicate the Nb depletion in the silicate mantle was produced before the Archean, with negligible Ti depletion \citep{nesbitt1979komatiites,arndt2003komatiites,munker2003evolution}. This observation cannot easily be accommodated with the formation of refractory rutile-bearing hidden eclogite reservoir  \citep{rudnick2000rutile,kamber2000role,nebel2010deep}, which might produce both Nb and Ti depletion in the accessible Earth \citep{schmidt2004dependence}. \bigskip


Experimental studies show that Nb becomes siderophile under reducing conditions ($ \log f_{\ce{O2}} \leq \mathrm{IW} - 2 $), whereas Ti remains lithophile \citep[\cref{fig:D-values}B;][]{wade2001earth,corgne2008metal,mann2009evidence,namur2016sulfur,cartier2020no,steenstra2020geochemical}. Thus, a metal-silicate differentiation under reduced conditions can selectively separate Nb from silicate melts, and leave Ti as oxide in the mantle. Therefore, the siderophile behavior of Nb under reduced conditions might be a most plausible scenario for the origin of the BSE's Nb depletion \citep{wade2001earth,munker2003evolution}. Recently, \citet{munker2017silicate} experimentally showed that Nb becomes less chalcophile when Fe/S of a sulfide melt increases. Thus, it is likely that the siderophile behavior of Nb in the Earth reflects the low S content of the Earth's core \citep{dreibus1996cosmochemical,mcdonough2014compositional}. \bigskip

\subsection{The Earth's building blocks}
\label{sec:building_block_Earth}

The fractionated RLE ratios in EC components provide insights into the nature of the Earth's building blocks, which are considered to be dominated by EC-like materials based on isotopic constraints \citep[e.g.,][]{javoy2010chemical,warren2011stable,dauphas2017isotopic,boyet2018enstatite}. Multiple observations suggest that the BSE has CI-like RLE ratios \citep[e.g., Al/Ti, Zr/Hf, Ca/Al and Sm/Nd, excepting possibly Nb/RLE;][]{mcdonough1995composition,munker2003evolution,stracke2012refractory,bouvier2008lu,bouvier2016primitive,burkhardt2016nucleosynthetic,willig2019earth,hasenstab2020evolution}. If the Earth's accretion was dominated by highly reduced, EC-like materials, their silicate fraction might have fractionated RLE ratios due to depletion of nominally lithophile, but moderately chalcophile refractory elements (e.g., Ti, Zr, Ca, REE)\add{, as we have observed in EC chondrules (Figs.~2-4)}. To produce the CI-like RLE ratios of the BSE, the reduced silicates and RLE-bearing sulfides should have been jointly incorporated into the accreting Earth. Subsequently, RLE hosted in these sulfides \add{from the impactors} must have been re-incorporated into the silicate mantle, which \change{requires}{is consistent with} a progressive oxidation of the Earth's interior \citep{corgne2008metal,mann2009evidence,schonbachler2010heterogeneous,rubie2011heterogeneous,rubie2015accretion}. An important implication of this scenario is that the Earth's \change{silicate and sulfide building blocks}{silicates and sulfides} should have originated in the same chemical reservoir which originally had a solar-like unfractionated RLE composition. In addition, these \change{disk materials}{silicates and sulfides} should have accreted to planetesimals before being physically separated in the protoplanetary disk. \bigskip

Alternatively, the distinct RLE composition of silicate fractions of EC and Earth might reflect distinct formation temperatures of their precursors. Importantly, the bulk Earth is depleted in S, but have chondritic Fe/Si value \citep{dreibus1996cosmochemical,mcdonough2014compositional}, indicating that its S depletion was established during a nebular condensation, rather than separation of sulfides from silicates by the physical sorting within the accretionary disk. Under highly reducing nebular conditions as recorded by EC, refractory sulfides such as oldhamite and niningerite condense at 1000--800 K, whereas troilite condenses at $ < $700 K as a major S-bearing phase \citep{hutson2000multi,pasek2005sulfur}. Therefore, RLE compositions of highly reduced silicates condensed before the \change{precipitation}{condensation} of these sulfides should not have been fractionated due to sulfide-silicate separation as recorded in EC chondrules. This scenario supports the condensation of Earth's precursors at higher temperatures as compared to the EC materials, which have been previously proposed based on the EC's elevated abundances of moderately volatile elements and low Mg/Si and RLE/Mg ratios of EC \citep{kerridge1979fractionation,larimer1979condensation,dauphas2015planetary,morbidelli2020subsolar,yoshizaki2020earth}. \bigskip

Our observations revealed distinct RLE compositions of the silicate fractions of EC and Earth, which further emphasize \change{the}{their} compositional distinctions \remove{between EC and Earth} \citep{palme1988moderately,mcdonough1995composition,yoshizaki2020earth}. The distinction between EC and Earth reflects an unrepresentative sampling of the solar system materials in our meteorite collections due to restricted source regions of meteoritic samples \citep{demeo2014solar,heck2017rare} and/or a lack of leftovers of the Earth's building blocks in the present-day solar system \citep{drake2002determining}. Although there is no meteorite group that matches the chemical and isotopic composition of Earth, EC and their components provide unique and useful guides to the origin, nature, and geochemical evolution of the Earth-forming materials \citep{dauphas2015planetary,morbidelli2020subsolar,yoshizaki2020earth}.



\subsection{Implications for the composition of Mercury}
\label{sec:mercury}

The RLE fractionation observed in the silicate fraction of EC (\cref{fig:chondrule_volatility_PP,fig:chondrule_volatility_others,fig:chondrule_RLEratio}) indicates that caution is needed when modeling RLE abundance of highly reduced bodies that have experienced an open-system metal-silicate separation (e.g., physical sorting before and during planetary accretion, collisional erosion of a mantle). The most notable example is Mercury, which is one of the most reduced bodies in our solar system and has an elevated bulk Fe/Si ratio that might reflect a selective removal of silicates and/or preferred accretion of metal/sulfide phases \citep{nittler2011major,ebel2017elusive,margot2018mercury}. \bigskip

The elemental abundance of the Mercury's surface was measured by the MErcury Surface, Space ENvironment, GEochemistry, and Ranging (MESSENGER) spacecraft, which provided important constraints on the chemical composition and interior structure of the planet \citep{nittler2011major,nittler2017chemical,nittler2020global,peplowski2011radioactive}. Recently, \citet{cartier2020no} showed that Al/Ti ratio in the bulk silicate Mercury is $ \sim $50\% higher than the CI value, based on the MESSENGER data. By assuming that the bulk Mercury has a chondritic Al/Ti value, the authors concluded that the Ti depletion in the Mercury's surface was produced by a S-bearing core formation under reducing condition, and estimated a thickness of a troilite layer at the core-mantle boundary. In contrast, chondrules and sulfides from EC show clear evidence for Al/Ti fractionation in a highly reduced nebular environment before planetary accretion and differentiation (\cref{fig:chondrule_RLEratio}). Therefore, the super-CI Al/Ti value of the Mercury's silicate shell might reflect both nebular and planetary RLE fractionation processes under reducing conditions. Importantly, if the planet preferentially accreted metal/sulfide phases \citep{wurm2013photophoretic,kruss2018seeding,kruss2020composition}, bulk Mercury can be enriched in moderately chalcophile RLE (e.g., Ti, REE, Nb, Ca) compared to Al, thereby invalidating the constant RLE ratio rule. Thus, the constant RLE rule should be used with caution, at least when estimating chemical compositions of highly reduced bodies. \bigskip

As we described in \cref{sec:building_block_Earth}, highly reduced silicates condensing from a cooling nebular gas before the \change{precipitation}{condensation} of RLE-bearing sulfides may not be affected by a significant RLE fractionation due to sulfide-silicate separation. The observations by the MESSENGER mission indicate that the surface materials of Mercury have less fractionated RLE ratios and lower abundances of moderately volatile elements as compared to the silicate fractions of EC \citep{nittler2011major,nittler2017chemical,nittler2020global,peplowski2011radioactive,evans2015chlorine,cartier2020no}. These observations might suggest that the building blocks of Mercury have condensed at higher temperatures than EC materials (i.e., before sulfide condensation), as we proposed for the Earth's precursor materials. \bigskip

\section{Conclusions}

Refractory lithophile elements (RLE) can be fractionated due to their different chalcophile affinities under highly reducing nebular conditions. Chondrules from enstatite chondrites (EC) record RLE fractionation due to a sulfide-silicate separation before and/or during chondrule formation events under a highly reduced nebular environments, documenting that RLE ratios \change{are not constant}{can be variable} among \add{silicates in} the planetary building blocks. In contrast, the bulk EC have CI-like RLE ratios, indicating a negligible physical sorting of silicates and sulfides before and during the accretion of EC parent bodies. Similarly, the Earth's building blocks might have not experienced the physical sorting of silicates and sulfides. Alternatively, they have condensed from a high-temperature nebular gas in which RLE-bearing sulfides were not stable. The accessible Earth's Nb depletion and its chondritic Ti/RLE values might reflect the moderately siderophile behavior of Nb during the Earth's accretion. The moderately chalcophile behavior of RLE under highly reducing conditions indicate that the bulk Mercury cannot yield CI-like RLE ratios if its large core/mantle ratio is originated in a selective removal of silicates.

\section*{Acknowledgments}

We thank Roberta Rudnick and Philip Piccoli for discussions and technical assistance, Issei Narita for technical assistance, Dominik Hezel for kindly providing the ChondriteDB spreadsheet. We appreciate Conel Alexander and Kevin Righter for their helpful comments. We greatly appreciate Herbert Palme, Andreas Stracke, Dominik Hezel, and an anonymous referee for their constructive reviews, which helped improve the manuscript. We thank the associate editor Stefan Weyer for his editorial effort. We are grateful to NASA/JSC for loan of ALH 84170, ALH 84206, ALH 85119, MAC 88136 and RBT 04143, National Institute of Polar Research for ALHA77295, and the Smithsonian Institute for Allende. This work was supported by Grant-in-Aid for JSPS Research Fellow (No.~JP18J20708), GP-EES Research Grant and DIARE Research Grant to TaY, JSPS KAKENHI Grant (No.~16H04081) to TeY, and NSF grant EAR1650365 to WFM.

\section*{Author contributions}

\textbf{Takashi Yoshizaki:} Conceptualization, Methodology, Validation, Formal analysis, Investigation, Resources, Data Curation, Writing--Original Draft, Visualization, Project administration, Funding acquisition.
\textbf{Richard D. Ash:} Methodology, Validation, Formal analysis, Investigation, Writing--Review \& Editing.
\textbf{Marc D. Lipella:} Formal analysis, Investigation, Data Curation.
\textbf{Tetsuya Yokoyama:} Writing--Review \& Editing, Funding acquisition.
\textbf{William F. McDonough:} Conceptualization, Methodology, Validation, Formal analysis, Writing--Review \& Editing, Funding acquisition.


\section*{Competing financial interests}

The authors declare no competing financial interests.

\section*{Supplementary materials}
Supplementary materials associated with this article can be found in the online version.

\clearpage

\begin{table}[p]
\centering
\caption{List of meteorite samples used in this study. Meteorite types and weathering grades are based on the online Meteoritical Bulletin Database (\url{www.lpi.usra.edu/meteor/}) and \citet{AM_newsletter_2008}. Abbreviations: NIPR--National Institute of Polar Research; SI--Smithsonian Institution.}
\begin{tabular}{cccc}
\toprule
Name &Type & Weathering grade & Sample source \\
\midrule
ALHA77295 & EH3  &        B       & NIPR   \\
ALH 84206 & EH3  &       A/B      & NASA/JSC \\
ALH 84170 & EH3  &        B          & NASA/JSC   \\
MAC 88136 & EL3  &       A     & NASA/JSC   \\
ALH 85119 & EL3  &       Be       & NASA/JSC   \\
RBT 04143  & CV3  &      B     & NASA/JSC  \\
Allende  & CV3  &     Fall    & SI   \\
\bottomrule
\end{tabular}%
\label{tab:sample_list}%
\end{table}%
\clearpage

\begin{table}[p]
\centering
\caption{Conditions of the LA-ICP-MS measurement.}
\label{tab:LA-ICP-MS_cnd}%
\begin{tabular}{ll}
\toprule
\multicolumn{2}{l}{Thermo Finnigan Element2 single-collector, sector field ICP-MS}  \\
\midrule
Forward power & 1265 W \\
HV    & 8 kV \\
Number of pre-scans & 1 \\
Active dead time & 18 ns \\
Cool gas flow & 16 L/min Ar \\
Auxiliary gas flow & 1.05 L/min Ar \\
Sample gas flow & 1.040 L/min Ar \\
Autosampler & ASX-100 (CETAC Technologies) \\
Torch & Quartz \\
Injector & Quartz \\
Skimmer cone & Al-alloy \\
Sampling cone & Al-alloy \\
Detection mode & Analogue or both \\
Dwell time & 20 s background acquisition, 40 s data acquisition \\
Samples/peak & 1 \\
Runs*passes & 100*1 \\
Search window & 150\% \\
Integration window & 80\% \\
Integration type & Intensities \\
Scan type & EScan \\
Oxide forming rate & \ce{^{238}U^{16}O}/\ce{^{238}U} $ < $ 2.0\% \\
\midrule
\multicolumn{2}{l}{Electro Scientific New Wave frequency-quintupled Nd:YAG laser (213 nm) system} \\
\midrule
Ablation pattern & Spot \\
Spot diameter & 15--200 $ \upmu $m \\
Repetition rate & 7 Hz \\
Energy density & 2--3 $ \mathrm{J/cm^{2}} $ \\
\bottomrule
\end{tabular}%
\end{table}%

\clearpage

\clearpage

\clearpage

\clearpage

\begin{figure}[h]
\centering
\includegraphics[width=1.0\linewidth]{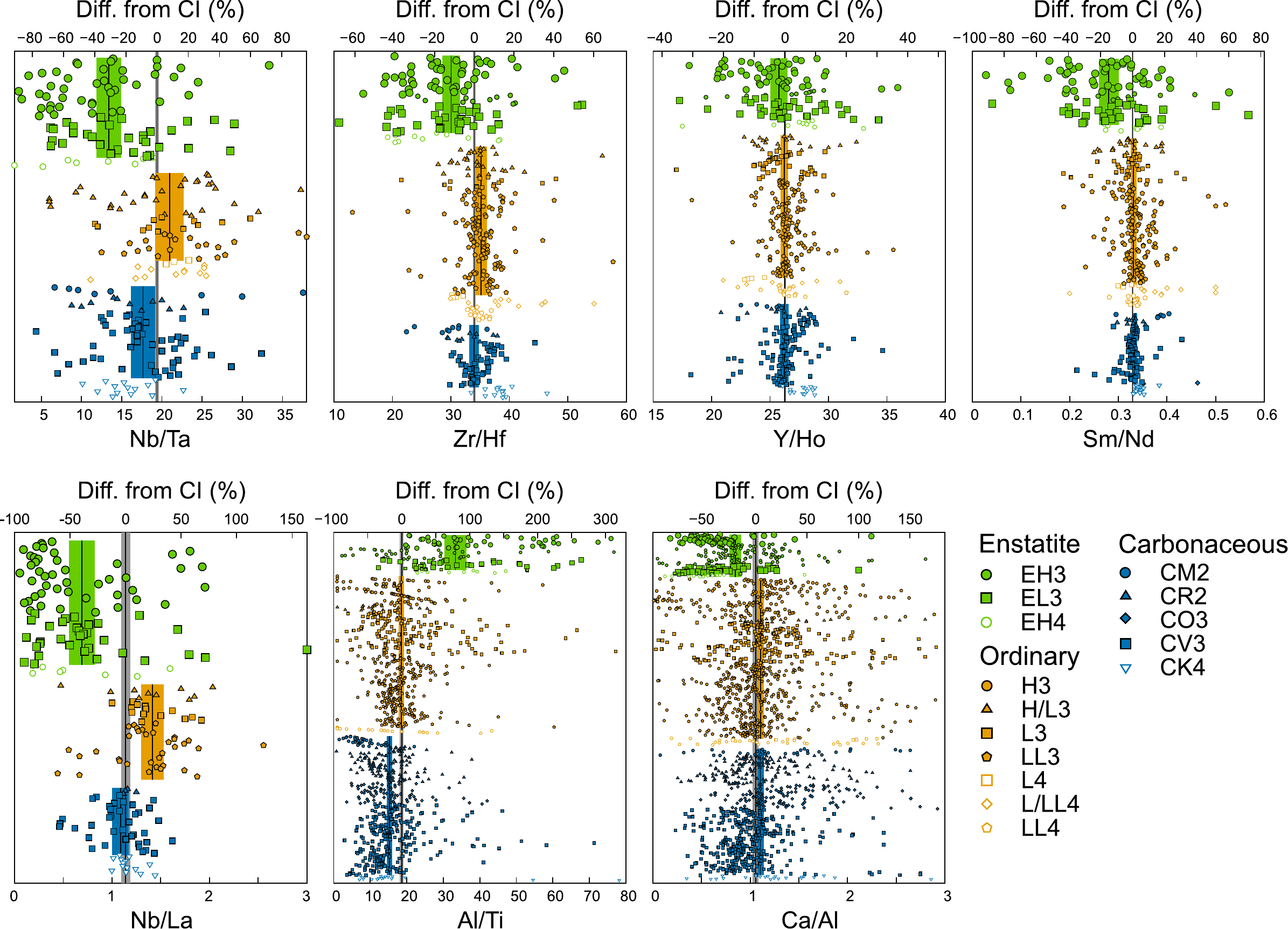}
\caption{Refractory lithophile element ratios (Nb/Ta, Zr/Hf, Y/Ho, Sm/Nd, Nb/La, Al/Ti, and Ca/Al) of bulk chondrules from unequilibrated type 3 enstatite (green), ordinary (orange) and carbonaceous (blue) chondrites. Our new data for type 3 EC chondrules are shown by large filled green circles (EH3) or squares (EL3). Colored boxes and inside solid lines represent mean $ \pm ~ 2\sigma _{\mathrm{m}} $ values for each chondrite groups. Gray box and inside solid line correspond mean $ \pm ~ 2\sigma _{\mathrm{m}} $ for CI chondrite \citep{barrat2012geochemistry,lodders2020solar}. Data for chondrules from metamorphosed type 4 samples are also shown for a comparison. Most literature data are obtained from the ChondriteDB Database \citep{hezel2018what}. A list of full data sources is provided as a supplementary material.}
\label{fig:chondrule_RLEratio}
\end{figure}
\clearpage

\begin{figure}[h]
\centering
\includegraphics[width=1.0\linewidth]{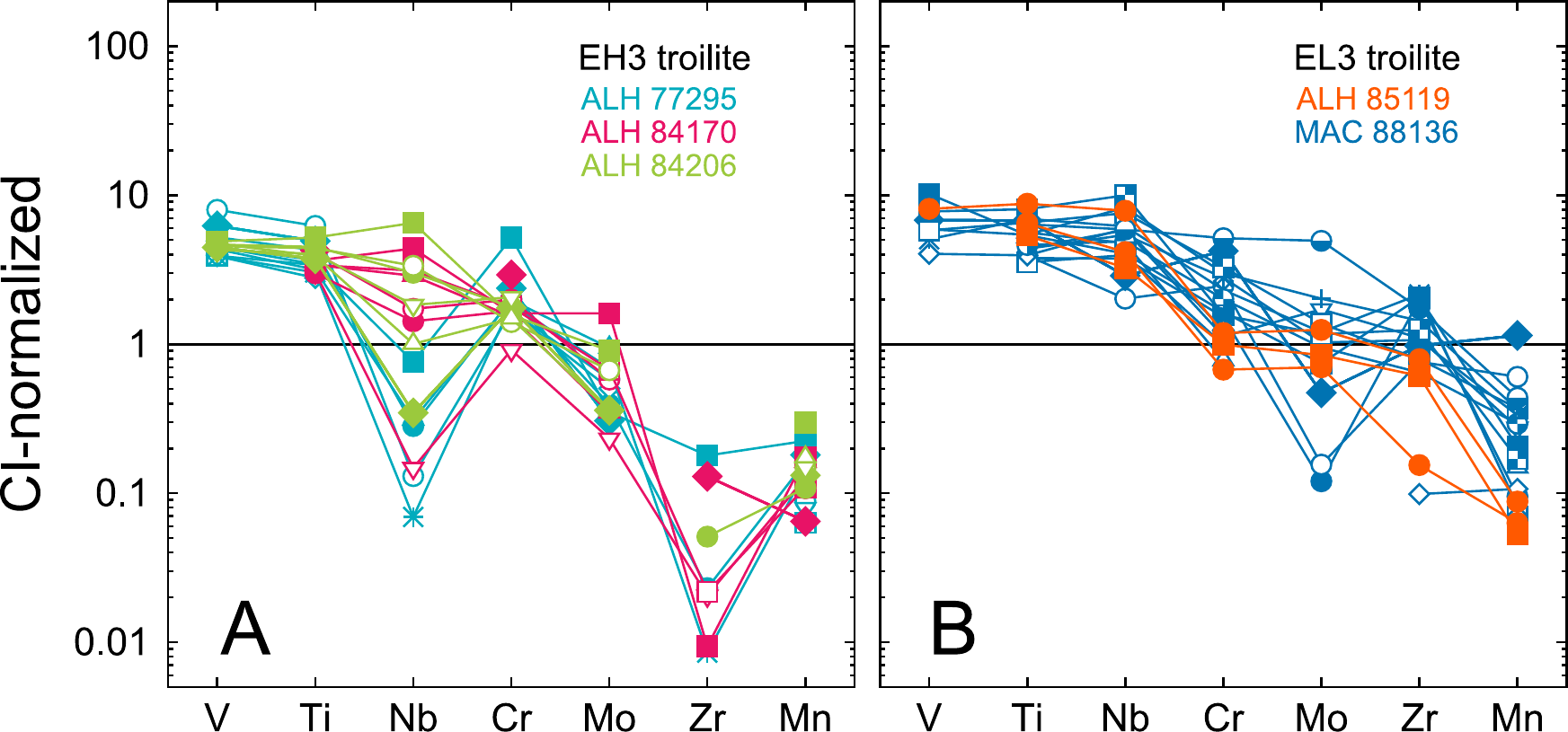}
\caption{Transition element composition of troilites (FeS) from unequilibrated EH and EL chondrite, normalized to CI chondrite composition \citep{lodders2020solar}.}
\label{fig:FeS_RLE}
\end{figure}
\clearpage

\begin{figure}[h]
\centering
\includegraphics[width=1.0\linewidth]{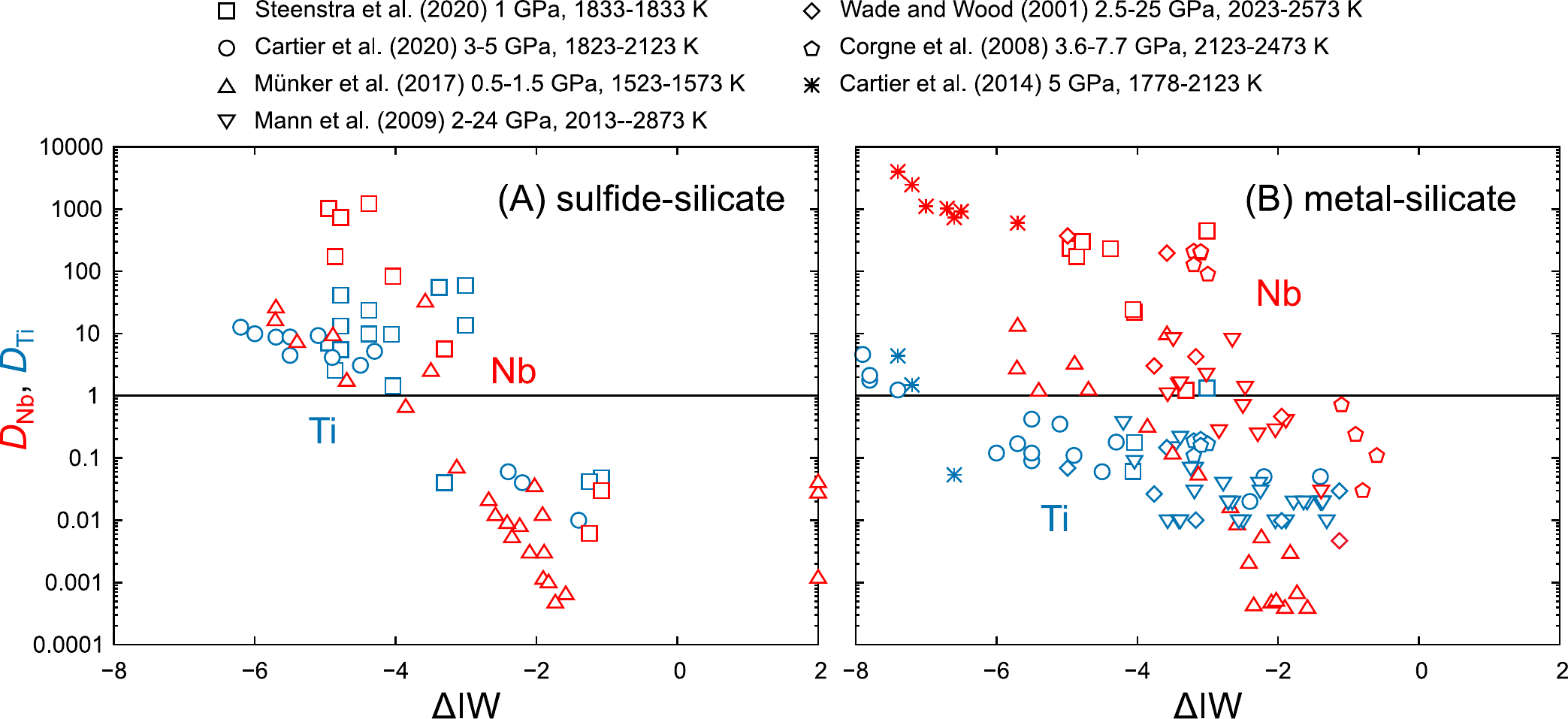}
\caption{(A) Sulfide-silicate and (B) metal-silicate partition coefficients of Nb (blue) and Ti (red) reported by previous studies \citep{wade2001earth,corgne2008metal,mann2009evidence,cartier2014redox,cartier2020no,munker2017silicate,steenstra2020geochemical}.}
\label{fig:D-values}
\end{figure}
\clearpage

\bibliographystyle{elsarticle-harv}
\bibliography{myrefs}

\putbib[myrefs]
\end{bibunit}

\newpage


\setcounter{page}{1}

\begin{appendices}

\counterwithin{figure}{section}
\counterwithin{table}{section}
\counterwithin{equation}{section}

\begin{bibunit}[elsarticle-harv]

\begin{center}
{\Large \textbf{Supplementary materials for \\
\vspace{1ex}
Variable refractory lithophile element compositions of planetary building blocks: insights from components of enstatite chondrites}} \\
\vspace{2ex}
Takashi Yoshizaki*$ ^1 $, Richard D. Ash$ ^2 $, Marc D. Lipella$ ^2 $, Tetsuya Yokoyama$ ^3 $, and \\ William F. McDonough$ ^{1,2,4} $
\vspace{1ex}

$ ^1 $Department of Earth Science, Graduate School of Science, Tohoku University, Sendai, Miyagi 980-8578, Japan \\
$ ^2 $Department of Geology, University of Maryland, College Park, MD 20742, USA \\
$ ^3 $Department of Earth and Planetary Sciences, Tokyo Institute of Technology, Ookayama, Tokyo 152-8851, Japan \\
$ ^4 $Research Center for Neutrino Science, Tohoku University, Sendai, Miyagi 980-8578, Japan \\
(*Corresponding author. E-mail: tky@dc.tohoku.ac.jp) \\
\vspace{1ex}

\today

\end{center}

\section{Supplementary tables}

Supplementary tables listed below are provided as Microsoft Excel spreadsheets.\\

\noindent Table A.1: Chemical composition of chondrules from primitive enstatite chondrites.

\noindent Table A.2: Chemical composition of troilite from primitive enstatite chondrites.

\noindent Table A.3: Chemical composition of oldhamite and niningerite from primitive enstatite chondrites.

\noindent Table A.4: Chemical composition of refractory inclusions from carbonaceous chondrites.

\noindent Table E.5: Summary of element ratios of chondrules from unequilibrated enstatite, ordinary, and carbonaceous chondrites.

\clearpage

\section{Supplementary text}

\subsection{Scanning electron microscope and electron microprobe analyses}
\label{sec:SM_SEM}

Mineralogy and petrology of polished sections of the enstatite chondrite samples (\cref{tab:sample_list}) were analyzed using Hitachi S-3400N scanning electron microscope (SEM) and JEOL JSM-7001F field-emission (FE) SEM at Tohoku University (TU). Semi-quantitative analyses of constituent minerals and X-ray mapping were performed using Oxford INCA energy-dispersive spectrometers (EDS) equipped with the SEM instruments, at an accelerating voltage of 15 kV and a beam current of 1.0--1.4 nA. \bigskip

Quantitative X-ray microanalysis of chondrules from enstatite chondrites and refractory inclusions from RBT 04143 was performed using JEOL JXA-8530F FE-type electron probe microanalyzer (EPMA) and JEOL JXA-8800 EPMA equipped with wavelength-dispersive X-ray spectrometers (WDS) at TU. A defocused electron beam (20--200 $\upmu$m in diameter) with an acceleration voltage of 15 kV and  beam current of 10 nA was used to determine abundances of 13 elements (Si, Ti, Al, Cr, Fe, Mn, Mg, Ca, Na, K, S, P and Ni) in bulk chondrules. The peak counting times were 10 s for Na; 20 s for Si, Al, Mg and Ca; and 40 s for Ti, Cr, Fe, Mn, K, P and Ni. To obtain bulk chemical composition, chondrules larger than 200 $\upmu$m in diameter were measured multiple (up to 11) times and the average value was calculated. Well-characterized natural and synthetic crystalline oxides and metals were used as standards. Matrix corrections were applied using the atomic number (Z), absorption (A), and fluorescence (F) (ZAF) correction method. The detection limits of measurements using the JXA-8530F FE-EPMA were 0.02 wt\% for \ce{CaO} and \ce{K2O}; 0.03 wt\% for \ce{SiO2}, \ce{MgO} and \ce{Na2O}; 0.04 wt\% for \ce{Al2O3}, \ce{Cr2O3}, \ce{FeO}, \ce{MnO} and \ce{P2O5}; 0.05 wt\% for \ce{NiO}; 0.07 wt\% for \ce{SO3}; and 0.16 wt\% for \ce{TiO2}. The detection limits of analyses by the JXA-8800 EPMA were 0.01 wt\% for \ce{CaO} and \ce{K2O}; 0.02 wt\% for \ce{Al2O3}, \ce{Na2O} and \ce{SO3}; 0.03 wt\% for \ce{MnO} and \ce{MgO}; 0.04 wt\% for \ce{SiO2}, \ce{TiO2}, \ce{Cr2O3} and \ce{FeO}; and 0.05 wt\% for \ce{P2O5} and \ce{NiO}. \bigskip

A focused electron beam ($ \sim $3 $\upmu$m in diameter) accelerated at 15 kV with a beam current of 10 nA was used to quantify 13 elements (Si, Ti, Al, Cr, Fe, Mn, Mg, Ca, Na, K, P, Ni and S) in sulfides using the JXA-8800 EPMA at TU. Peak counting time was 20 s for all elements. Other analytical conditions were similar to those of chondrule measurements. Detection limits were 0.01 wt\% for Si, Al, Mg, Ca, K, P and S; 0.02 wt\% for Ti and Na; 0.03 wt\% for Cr, Fe and Mn; and 0.04 wt\% for Ni. \bigskip

Major element abundance in refractory inclusions from Allende was determined using and JEOL JXA-8900R EPMA at University of Maryland. Analyses were performed in a semi-quantitative mode. Wavelength scans were performing with the following conditions: 15 kV accelerating potential, 150 nA cup current, and a 20 $ \upmu $m beam diameter. Raw X-ray intensities were corrected using the ZAF algorithm. \bigskip

\subsection{Comparison of the bulk chemical composition of EC chondrules with previous data}
\label{SM_EC_data_comparison}

We recognize that the bulk composition of individual chondrules derived by surface analytical methods (i.e., broadened or scanning beam measurements on thin or thick sections) can be less accurate than those determined by a whole-rock measurement of mechanically separated chondrules. In particular, a biased sampling of phases in chondrules can provide RLE data that are affected by melt-crystal elemental fractionations. Thus, the ratios of elements with different incompatibilities (e.g., Al/Ti, Ca/Al, Nb/La, Sm/Nd) can be sensitive to the biased surface sampling, whereas those of the geochemical twins (Nb/Ta, Zr/Hf, and Y/Ho) might remain to represent the bulk chondrule compositions. \bigskip

\begin{figure}[h]
\centering
\includegraphics[width=.9\linewidth]{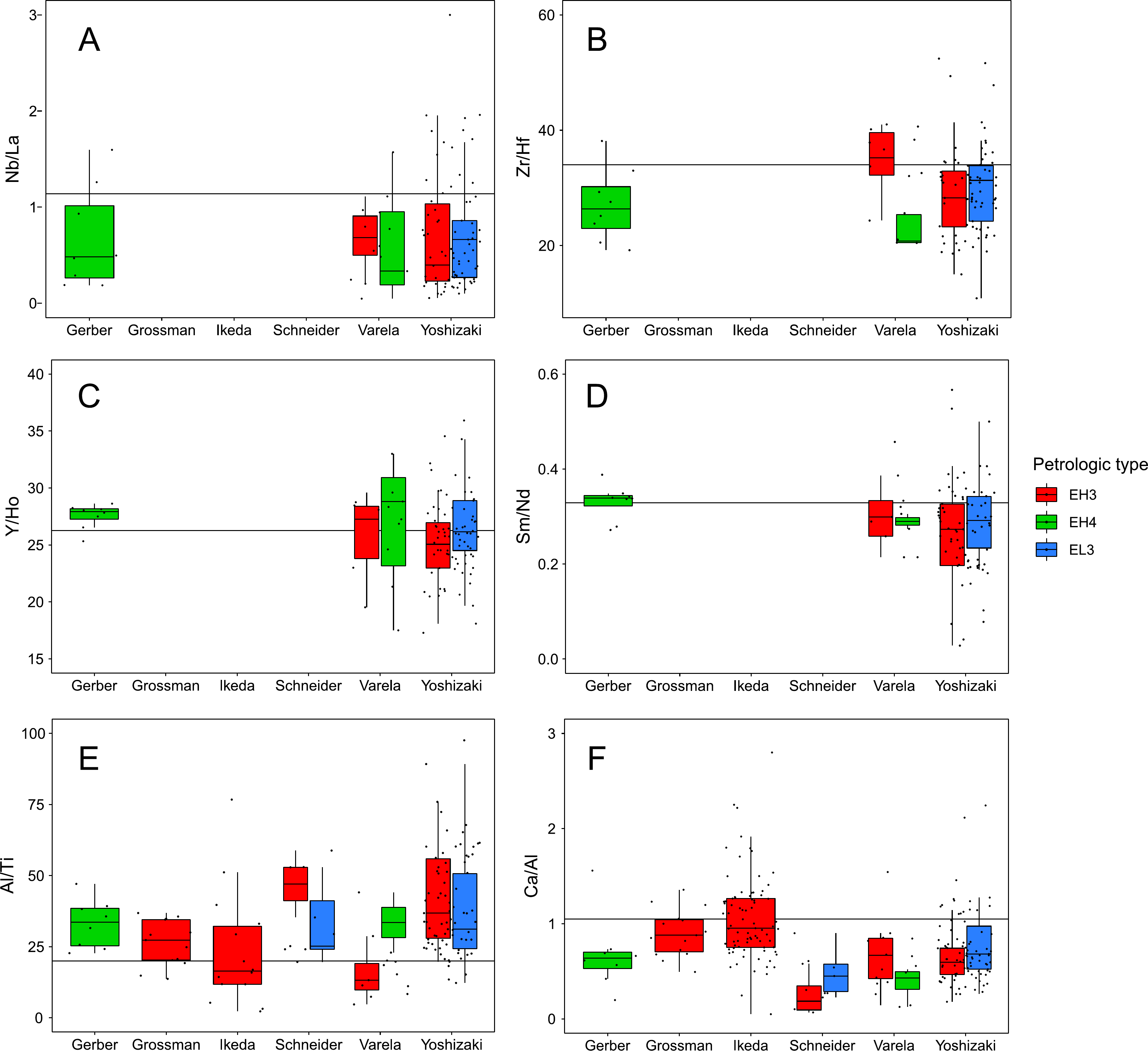}
\caption{Ratios of refractory lithophile elements in the bulk chondrules from EH3 (red), EH4 (green), and EL3 (blue) chondrites \citep[this study;][]{ikeda1983major,grossman1985chondrules,schneider2002properties,gerber2012chondrule,varela2015nonporphyritic}. Horizontal lines represent the CI ratio \citep{lodders2020solar}.}
\label{fig:EC_comparison_all_ratios}
\end{figure}

Recently. \citet{gerber2012chondrule} obtained bulk compositions of eight chondrules from type 4 EH Indarch, by analyzing solutions of mechanically isolated individual chondrules using ICP-MS and ICP-OES. Although the number of data reported by \citet{gerber2012chondrule} is limited and all of them are obtained from a metamorphosed type 4 EH, they can be used to investigate representativeness of the bulk chondrule data obtained by the surface analytical methods. Our data of Nb/La, Zr/Hf, Al/Ti, and Ca/Al in EC chondrules are consistent with those from \citet{gerber2012chondrule} (\cref{fig:EC_comparison_all_ratios}). EH4 chondrules from \citet{gerber2012chondrule} show slightly higher Y/Ho and Sm/Nd ratios than our data for EH3 and EL3 chondrules, which might reflect modifications of these ratios due to distinct behavior of these elemental twins during thermal processing in the EC parent bodies \citep{barrat2014lithophile}. \bigskip

\citet{varela2015nonporphyritic} obtained bulk chemical compositions of non-porphyritic chondrules from EH3 (Sahara 97158) and EH4 (Indarch) using the electron microprobe and LA-ICP-MS techniques. In general, their data are also consistent with data obtained by this study and \citet{gerber2012chondrule} (\cref{fig:EC_comparison_all_ratios}). On the other hand, \citet{varela2015nonporphyritic} reported elevated Zr/Hf and low Al/Ti ratios of EH3 chondrule as compared to our results, whereas in EH4 these ratios are consistent with the results of \citet{gerber2012chondrule}. Ratios of major RLE in type 3 EC obtained in this study are also consistent with electron microprobe data from \citet{grossman1985chondrules} (Qingzhen (EH3)) and \citet{schneider2002properties} (Allan Hills (ALH) 85119 (EL3); MacAlpine Hills (MAC) 88180 (EL3), and Pecora Escarpment (PCA) 91020 (EL3), PCA 91238)). \bigskip

\begin{figure}[h]
\centering
\includegraphics[width=.9\linewidth]{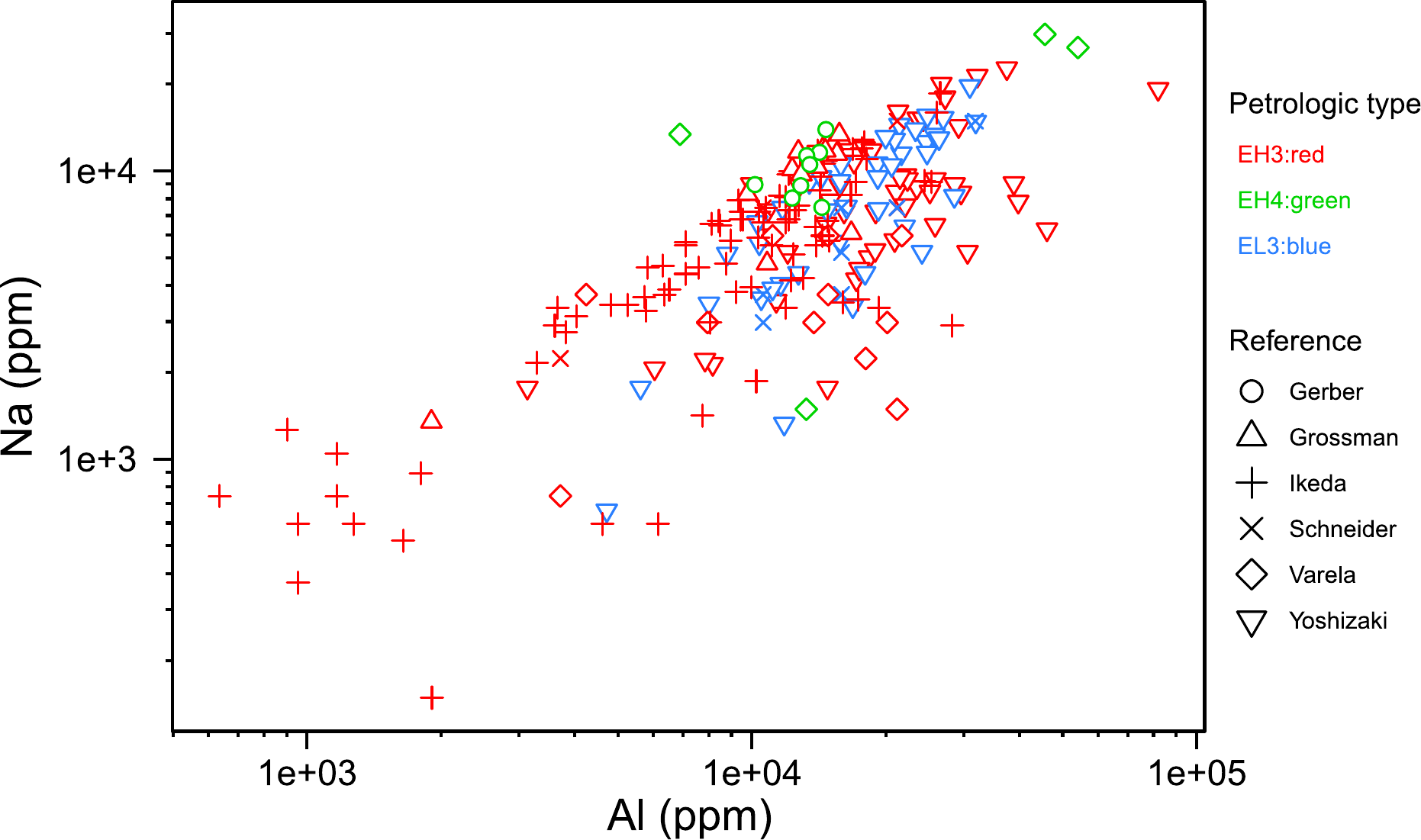}
\caption{Abundances of Na and Al in bulk chondrules from EH3 (red), EH4 (green), and EL3 (blue) chondrites \citep[this study;][]{ikeda1983major,grossman1985chondrules,schneider2002properties,gerber2012chondrule,varela2015nonporphyritic}.}
\label{fig:EC_comparison_all_NaAl}
\end{figure}

In contrast, electron microprobe data of bulk chondrules from Yamato 691 (EH3) obtained by \citet{ikeda1983major} show wide spread in Al/Ti and Ca/Al ratios, with mean values overlapping the CI ratios (\cref{fig:EC_comparison_all_ratios}). Since the data reported by \citet{ikeda1983major} show lower Na and Al abundances compared to other studies
\citep[this study;][; \cref{fig:EC_comparison_all_NaAl}]{grossman1985chondrules,gerber2012chondrule,varela2015nonporphyritic}, it is likely that their data are derived from glass-poor regions of the EC chondrules. \citet{ikeda1983major} noted that they avoided a beam overlap onto metals and sulfides in chondrules, which might also have led exclusion of a glassy mesostasis in which these opaque phases commonly enclosed \citep{weisberg2012unequilibrated,lehner2013formation,piani2016magmatic,jacquet2018chondrules}. We also avoided a beam overlap onto these opaque phases in our measurements, in which high spatial resolution images from optical and electron microscopes helped us to measure compositions of glassy areas without hitting sulfides. Alternatively, these discrepancies might represent unique compositional characteristics of chondrules from Yamato 691. Otherwise, our measurements might have overestimated the Al abundance, whereas our bulk Al/Ti and Ca/Al data are consistent with those from \citet{gerber2012chondrule}. Thus, we consider that the consistency of our bulk chondrule data with those of \citet{gerber2012chondrule} support representative sampling of bulk chondrule minerals in our measurements. \bigskip

\subsection{Composition of troilites from ALHA77295}
\label{SM_ALHA77295_FeS}

Troilites from ALH 84170 and ALH 84206 (EH3) have relatively uniform compositions, particularly for V through Mo (\cref{fig:FeS_RLE}). Likewise, most of the troilites from ALHA77295 (EH3), show comparable abundance patterns, but are depleted in Nb (0.1--0.8 $ \times $ CI) as compared to other EH3 samples. However, in three of the analyzed troilites from ALHA77295 (i.e., $ \sim $5$\%$ of the total troilite signal for ALHA77295) we encountered marked increases in Nb $ + $ Ta $ \pm $ Mo $ \pm $ W signals during the laser ablation analyses. These compositionally heterogeneous domains are of the order of 5 cubic microns and typically these inclusions had marked enrichment in $ \sim$500 $\upmu$g/g Nb, nearly chondritic Nb/Ta, and order ppm levels of Mo and W. Such inclusions were not encountered during analysis of troilite from any of the other EH3 and EL3 chondrites. No significant signal change was detected for the major elements when we encounter the heterogeneous domains. \bigskip

We found that the Nb concentration of bulk rock and of the acid leachate of ALHA77295 were similar to those of other unequilibrated EH chondrites \citep[this study;][]{barrat2014lithophile}. In ALHA77295 it is likely that some small fraction of the Nb and Ta and to a lesser extent Mo and W are hosted in these micron scale inclusions in troilite. We examined the post-ablation sampling area using FE-SEM/EDS, and found no additional metallic/sulfide inclusions. Therefore, the abundances of these elements in troilite from ALHA77295 reported in \cref{fig:FeS_RLE} should be considered as lower limits of their true concentrations. \bigskip

\subsection{Mineralogy and trace element chemistry of refractory inclusions from CV chondrites}
\label{sec:SM_CAI}

Here we provide a brief description of mineralogy and chemistry of refractory inclusions (three CAI and six AOA) from CV chondrites analyzed in this study (\cref{fig:CAI_AOA_volatility} and Table A.4). AOA are one of the most common type of refractory inclusions in carbonaceous chondrites \citep[e.g.,][]{,krot2014classification}. They are aggregates of fine-grained forsterite, Fe, Ni-metal and a variety of refractory minerals (e.g., Al-rich diopside, anorthite, spinel) \citep[e.g.,][]{,krot2004amoeboid}. Geochemical investigation of AOA indicate that these inclusions are nebular condensates that have no or minor degree of melting after their formation \citep[e.g.,][]{,krot2004amoeboid}, and thus they are considered to record fractionation of refractory elements occurred in the solar nebula \citep[e.g.,][]{,ruzicka2012amoeboid}.\bigskip

R7C-01 is a fragment of an irregularly shaped CAI that consists of fine-grained spinel and Al, Ti-rich diopside and a small amount of anorthite and surrounded by an Al-rich diopside rim. Anorthite is partially replaced by a secondary nepheline. The bulk CAI shows almost flat REE pattern ($\sim $ 40 $ \times $ CI) with negative anomalies in Eu and Yb and a small negative Ce anomaly.\bigskip

Based on mineralogy and bulk major element abundance, we classified AOA from RBT 04143 into two groups: Ca, Al-rich AOA and olivine-rich AOA. Ca, Al-rich AOA contain abundant nodules of Ca, Al-rich minerals that commonly occur in CAI \citep[e.g., Al-diopside, anorthite, spinel;][]{,brearley1998chondritic,macpherson2014calcium} with high bulk abundance of major RLE (e.g., Ca, Al, Ti; Table A.4). In contrast, such refractory minerals are nearly absent in olivine-rich AOA, which consist of abundant olivine and minor amounts of Fe, Ni-metal and Al-diopside.\bigskip

R1A-03 is an irregularly-shaped Ca, Al-rich AOA that is composed of anorthite-rich nodules with minor Al, Ti-rich diopside and spinel, and forsterite. This inclusion has a nearly flat REE pattern ($\sim $ 5 $ \times $ CI) with positive Ce and Eu anomalies. R4A-45 consists of a subrounded nodule that consists of a core of spinel + Al, Ti-rich diopside and Al-diopside rim, which is surrounded by a compact-textured forsterite thereafter. Spinel-rich domain of this inclusion is characterized by highly fractionated REE pattern with a higher light REE abundance (Ce, Pr and Nd; $\sim $ 20 $ \times $ CI)  than heavy REE (Gd, Tb, Dy, Ho, Er and Lu; $\sim $ 10 $ \times $ CI), and negative Eu and positive Sm, Tm and Yb anomalies. Two spot data were obtained from olivine-rich region of this inclusion. One measurement shows highly fractionated REE pattern as observed in the spinel-rich domain of this inclusion with much lower REE abundance ($\sim $ 5 $ \times $ CI and $\sim $ 2 $ \times $ CI for LREE and HREE, respectively). Another area of the olivine-rich domain shows nearly flat REE pattern ($\sim $ 5 $ \times $ CI) with a negative Eu anomaly. \bigskip

Four AOA from RBT 04143 are classified as olivine-rich AOA. Typical CAI minerals (e.g., Al, Ti-rich diopside, spinel) are nearly absent in these inclusions. R3A-18 is irregularly-shaped, olivine-rich inclusion with a compact texture with low REE abundance. R1A-04 is a fine-grained irregular AOA which consists of forsterite-rich core and a mantle of forsterite + Fe-rich olivine + Fe, Ni-metal. The core of this inclusion is characterized by nearly flat REE pattern ($\sim $ 2 $ \times $ CI). R1A-01 is irregularly shaped AOA that is dominated by forsterite with a minor amount of Fe, Ni-metal. The REE pattern of this inclusion is flat ($\sim $ 4 $ \times $ CI) with negative Eu anomaly. R1A-06 is irregularly-shaped AOA that consists of Fe, Ni-metal-rich core, forsterite-rich mantle with minor Al-diopside, and a rim of forsterite + Fe, Ni-metal. This inclusion shows flat REE pattern ($\sim $ 1--2 $ \times $ CI). \bigskip

For Allende CAI, we determined abundances of major elements and HFSE (Table A.4). 	The Allende CAI 3529-63-RR1 is a brecciated CAI with abundant spinel and anorthite. In this inclusion, anorthite occurs replacing primary melilite and minor Al, Ti-rich diopside is also identified. 3529-61-RR1 is a melilite-rich CAI with a well-developed rim sequence. Ca, Fe-rich pyroxene and sodalite is abundant in this inclusion, reflecting significant alteration of this inclusion in the Allende parent body. \bigskip

\begin{table}[p]
\centering
\caption{Long-term precision of LA-ICP-MS analyses. Abbreviations: CC--carbonaceous chondrites; EC--enstatite chondrites.}
\label{tab:precision}%
\begin{tabular}{ccccccccccc}
\toprule
Element & Mass & RSD\% &       & Element  & Mass & RSD\% &       & Element & Mass & RSD\% \\
\midrule
\multicolumn{11}{l}{NIST 610 (for EC chondrules and CC refractory inclusions, $ n=28 $)} \\
Ca    & 43    & 4     &       & Ce & 140 & 3 &       & Ho & 165 & 5 \\
Sc    & 45    & 4     &       & Pr & 141 & 3 &       & Er & 166 & 4 \\
Ti    & 47    & 5     &       & Nd & 146 & 3 &       & Tm & 169 & 5 \\
V     & 51    & 5     &       & Sm & 147 & 4 &       & Yb & 172 & 4 \\
Y     & 89    & 4     &       & Eu & 151 & 4 &       & Lu & 175 & 3 \\
Zr    & 90    & 4     &       & Gd & 157 & 4 &       & Hf & 178 & 2 \\
Nb    & 93    & 4     &       & Tb & 159 & 6 &       & Ta & 181 & 3 \\
Ba    & 137   & 5     &       & Dy & 163 & 5 &       & W & 182 & 4 \\
La    & 139   & 3     &       &       &       &       &       &       &       &  \\
&&&&&&&&&&\\
\multicolumn{11}{l}{NIST 610 (for EC sulfides, $ n=32 $)} \\
    Mg    & 25    & 3     &       & Ga    & 69    & 2     &       & Sm    & 147   & 3 \\
    Al    & 27    & 4     &       & Sr    & 88    & 3     &       & Eu    & 153   & 4 \\
    Si    & 29    & 3     &       & Y     & 89    & 4     &       & Gd    & 157   & 4 \\
    P     & 31    & 3     &       & Zr    & 90    & 3     &       & Ho    & 165   & 4 \\
    Ca    & 43    & 3     &       & Nb    & 93    & 3     &       & Tm    & 169   & 4 \\
    Sc    & 45    & 3     &       & Mo    & 95    & 4     &       & Yb    & 172   & 4 \\
    Ti    & 47    & 4     &       & Cs    & 133   & 4     &       & Lu    & 175   & 3 \\
    V     & 51    & 3     &       & La    & 139   & 4     &       & Hf    & 178   & 3 \\
    Cr    & 53    & 3     &       & Ce    & 140   & 4     &       & Ta    & 181   & 3 \\
    Mn    & 55    & 3     &       & Nd    & 146   & 4     &       & W     & 182   & 4 \\
    Zn    & 66    & 2     &       &       &       &       &       &       &       &  \\
&&&&&&&&&&\\
\multicolumn{11}{l}{Filomena (for EC sulfides, $ n=32 $)} \\
Ni & 60 & 10 &&&&&&&&\\
Cu & 63 & 8 &&&&&&&&\\
\bottomrule
\end{tabular}%
\end{table}%
\clearpage

\clearpage

\clearpage

\begin{figure}[h]
\centering
\includegraphics[width=.9\linewidth]{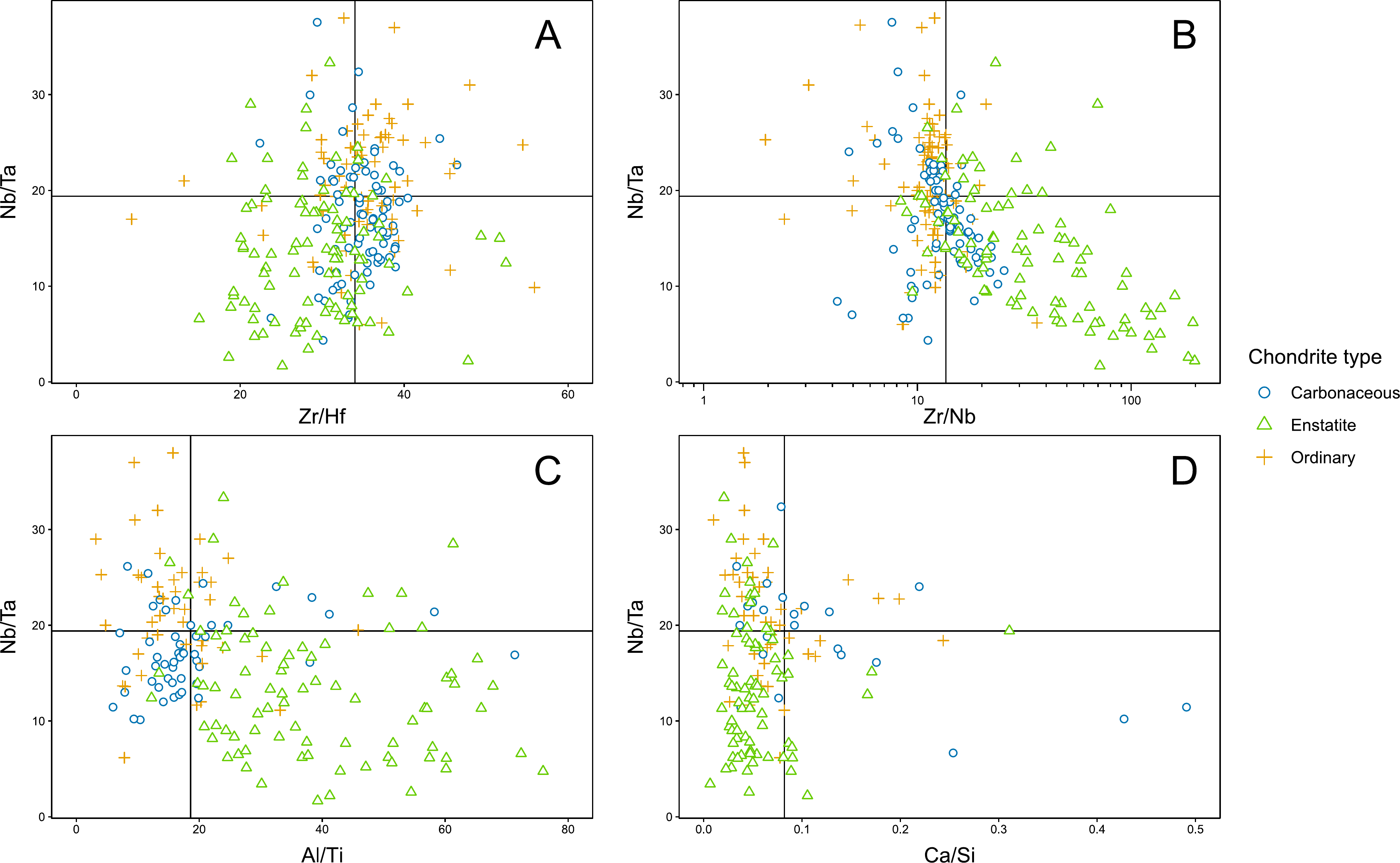}
\caption{Lithophile element ratios vs Nb/Ta in chondrules from enstatite (green), ordinary (orange), and carbonaceous (blue) chondrites. Data sources are similar to those of \cref{fig:chondrule_RLEratio}.}
\label{fig:chondrules_ratios_x-y}
\end{figure}
\clearpage

\begin{figure}[h]
\centering
\includegraphics[width=.9\linewidth]{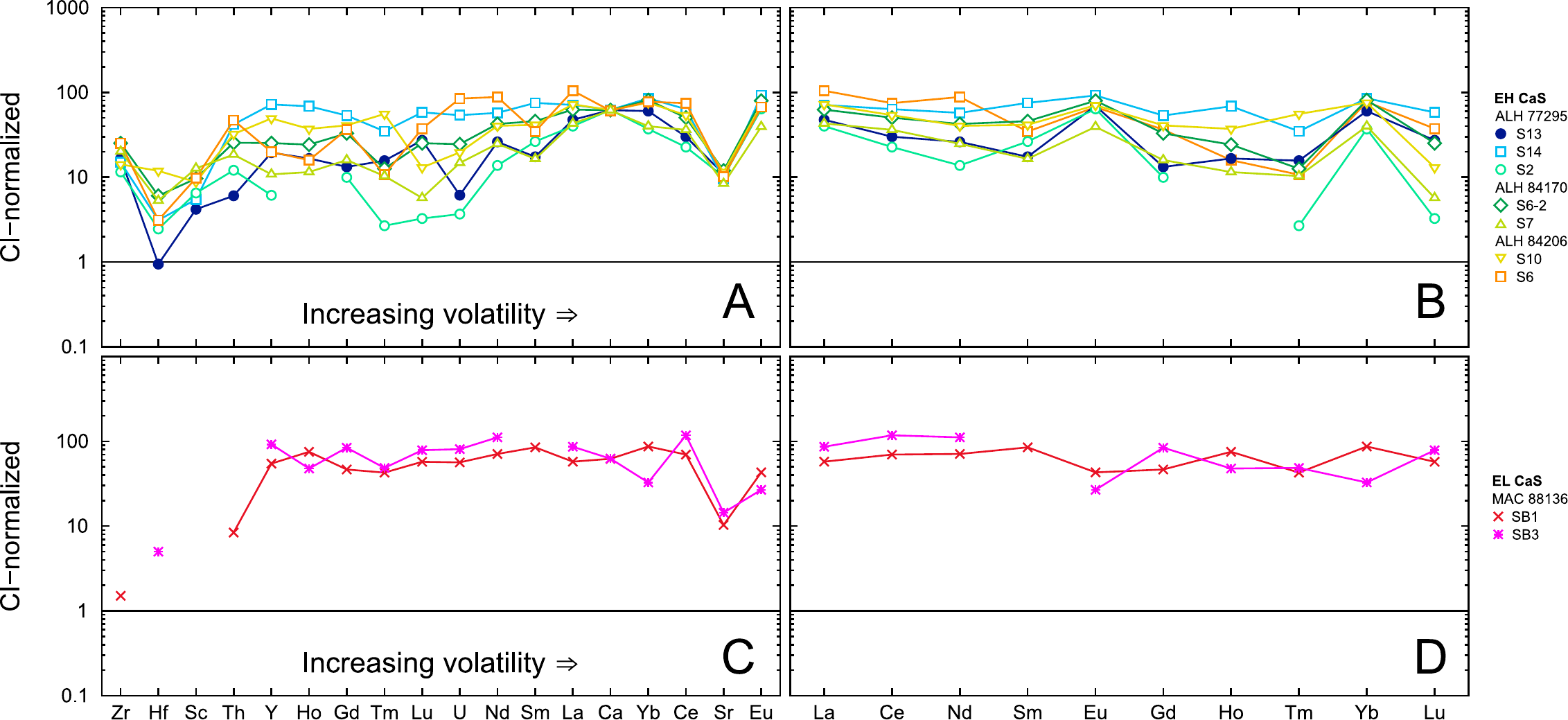}
\caption{CI-normalized \citep{lodders2020solar} abundances of lithophile elements in oldhamites (CaS) from EH (A,B) and EL (C,D) chondrites. In (A) and (B), elements are arranged in the order of increasing volatility \citep{lodders2003solar}.}
\label{fig:CaS}
\end{figure}
\clearpage

\begin{figure}[h]
\centering
\includegraphics[width=.9\linewidth]{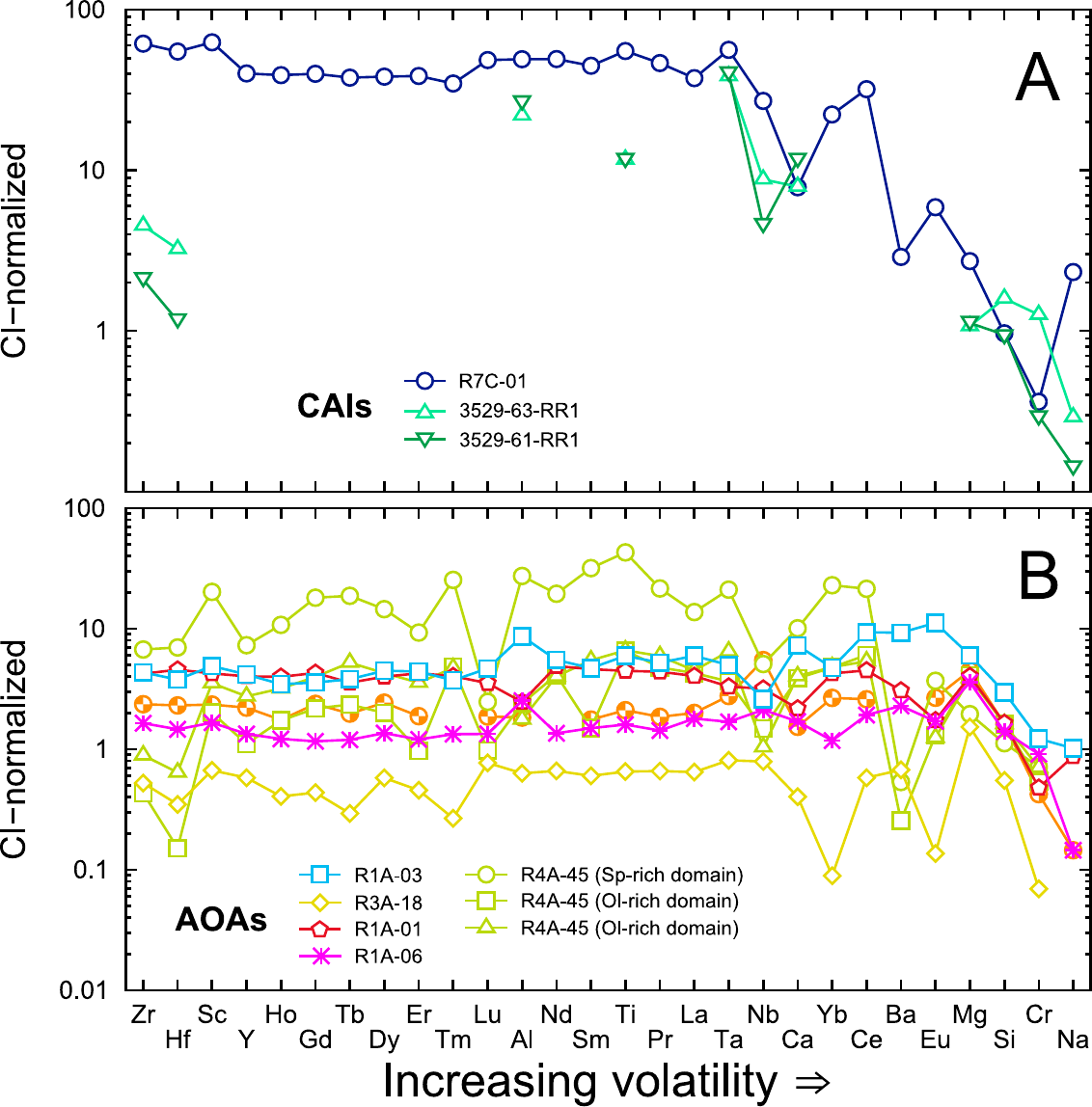}
\caption{CI-normalized \citep{lodders2020solar} abundances of lithophile elements in (A) Ca, Al-rich inclusions (CAI) and (B) amoeboid olivine aggregates (AOA) from unequilibrated carbonaceous chondrites. Abbreviations: Ol--olivine; Sp--spinel.}
\label{fig:CAI_AOA_volatility}
\end{figure}
\clearpage

\begin{figure}[h]
\centering
\includegraphics[width=.8\linewidth]{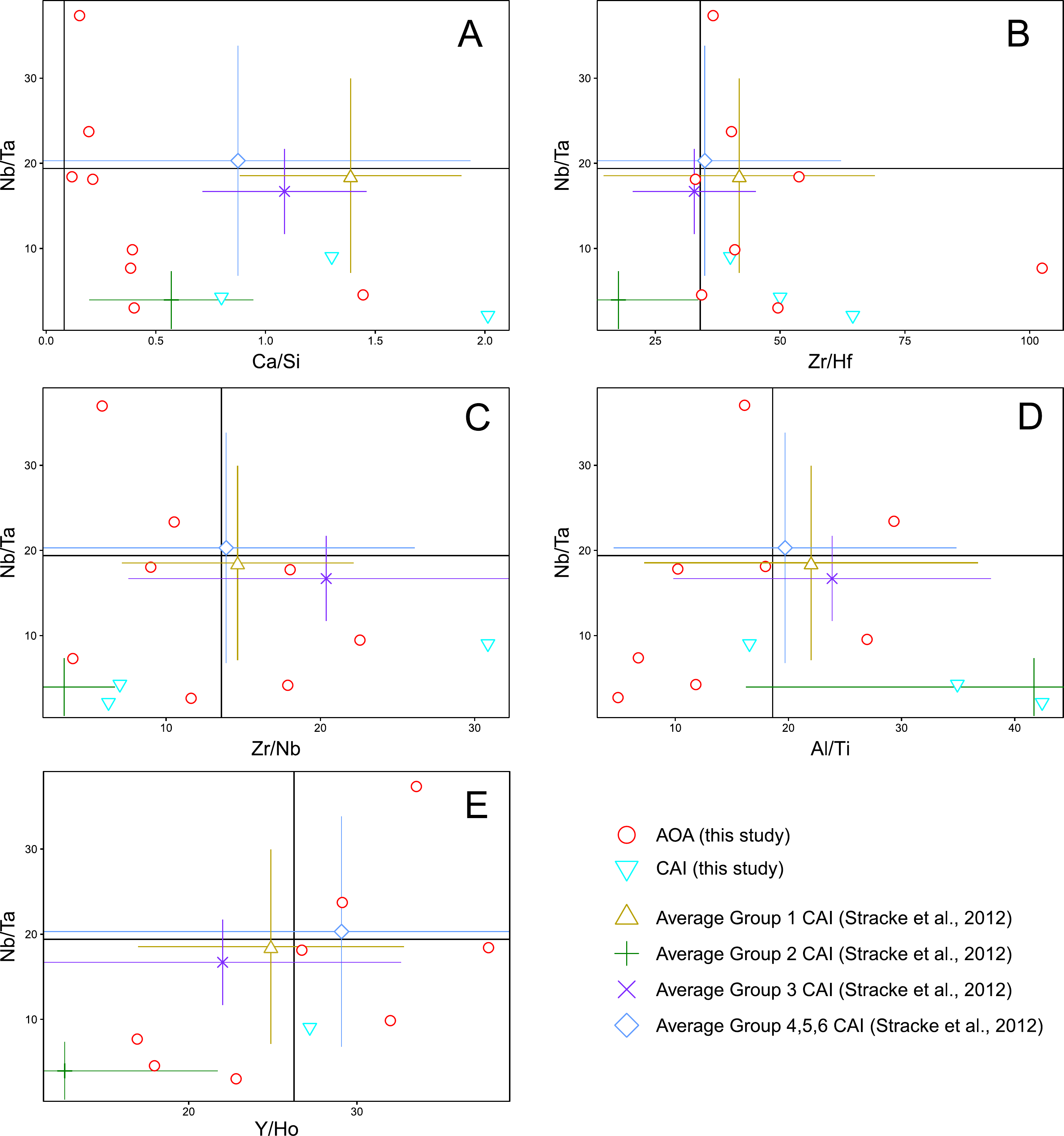}
\caption{Lithophile element ratios of bulk refractory inclusions (CAI and AOA) from CV chondrites obtained in this study. Also shown are average compositions with 1 standard deviations of each group of CAI \citep{stracke2012refractory}.}
\label{fig:rbt_summary}
\end{figure}
\clearpage

\renewcommand{\refname}{Supplementary references}
\putbib[myrefs]
\end{bibunit}

\clearpage

\section{Sources of chondrule data}
\label{sec:SM_data_sources}

Here we provide a list of sources of literature chondrule data used to produce \cref{fig:chondrule_RLEratio}. Most of these literature data are obtained through the online database MetBase (\url{https://metbase.org/}).


\renewcommand{\refname}{}

%
%
%
%
%
%
%
%

\end{appendices}

\end{document}